\documentclass[journal,twoside]{IEEEtran}

\usepackage{amsthm}
\usepackage{mathtools}
\usepackage{amsmath}
\usepackage{stmaryrd}
\usepackage{amsfonts}
\usepackage{dsfont}
\usepackage{amssymb}
\newtheorem{mydef}{Definition}
\newtheorem{mytheo}{Theorem}
\newtheorem{Lemma}{Lemma}

\newtheorem{Prop}{Proposition}
\newtheorem{Remark}{Remark}
\newtheorem{Coro}{Corollary}
\usepackage[latin1]{inputenc}
\usepackage{tikz}
\usetikzlibrary{shapes,arrows}
\usetikzlibrary{calc}
\usepackage{graphicx}
\usepackage{adjustbox}
\usepackage{citesort}
\usepackage{bbm}
\usepackage{color}
\usepackage{manfnt}
\usepackage{float}
\usepackage{pgfplots}
\pgfplotsset{compat=newest} 

\tikzstyle{block1} = [rectangle, draw, text width=1.4cm, text centered, rounded corners, minimum height=0.4cm]
\tikzstyle{block11} = [rectangle, draw, text width=1.9cm, text centered, rounded corners, minimum height=0.4cm]
\tikzstyle{block2} = [rectangle, draw, text width=1.5cm, text centered, rounded corners, minimum height=2cm]
\tikzstyle{block3} = [rectangle, draw , text width=1.5cm, text centered, rounded corners, minimum height=3.5cm]
\tikzstyle{block4} = [rectangle, draw, text width=2.1cm, text centered, rounded corners, minimum height=0.7cm]
\tikzstyle{block5} = [rectangle, draw, text width=0.25cm, text centered, minimum height=0.25cm]
\tikzstyle{dot} = [rectangle, text width=0.35cm, minimum height=0.5cm]
\tikzstyle{dot1} = [rectangle, text width=0.0cm, minimum height=0cm]
\tikzstyle{line} = [draw, -latex']

\newcommand{\sq}[2][0]{
  \mbox{$\medmuskip=#1mu\displaystyle#2$}%
}

\begin{document}

\title{On The Capacity of Broadcast Channels With Degraded Message Sets and Message Cognition Under Different Secrecy Constraints}

\author{Ahmed S. Mansour, Rafael F. Schaefer,~\IEEEmembership{Member,~IEEE}, and Holger Boche,~\IEEEmembership{Fellow,~IEEE}
	\thanks{This work was presented in part at IEEE-SPAWC, Tronto, Canada, June 2014 \cite{Mansour} and at IEEE-ITW, Hobart, Tasmania, Australia, November 2014 \cite{MansourITW}.}
	\thanks{Ahmed S. Mansour and Holger Boche are with the Lehrstuhl f\"ur Theoretische Informationstechnik, Technische Universit\"at M\"unchen, 80290 M\"unchen, Germany (e-mail:ahmed.mansour@tum.de; boche@tum.de). Rafael F. Schaefer is with the Information Theory and Applications Group,  Technische Universit\"at Berlin, 10587 Berlin, Germany (email: rafael.schaefer@tu-berlin.de).}
	\thanks{This work of R. F. Schaefer was supported by the German Research Foundation (DFG) under Grant WY 151/2-1.}
}

\markboth{Submitted For Publication}{BC With Degraded Message Sets and Message Cognition}

\maketitle
\vspace{-1.1cm}
\begin{abstract}
This paper considers a three-receiver broadcast channel with degraded message sets and message cognition. The model consists of a common message for all three receivers, a private common message for only two receivers and two additional private messages for these two receivers, such that each receiver is only interested in one message, while being fully cognizant of the other one. First, this model is investigated without any secrecy constraints, where the capacity region is established, showing that the straightforward extension of the K\"{o}rner and Marton inner bound to the investigated scenario is optimal. In particular, this agrees with Nair and Wang's result, which states that the idea of indirect decoding -- introduced to improve the K\"{o}rner and Marton inner bound -- does not provide a better region for this scenario. Further, some secrecy constraints are introduced by letting the private messages to be confidential ones. Two different secrecy criteria are considered: joint secrecy and individual secrecy. For both criteria, a general achievable rate region is provided. Moreover, the joint and individual secrecy capacity regions are established, if the two legitimate receivers are more capable than the eavesdropper. The established capacity regions indicate that the individual secrecy criterion can provide a larger capacity region as compared to the joint one, because each cognizant message can be used as a secret key for the other individual message. Further, the joint secrecy capacity is established for a more general class of more capable channels, where only one of the two legitimate receivers is more capable than the eavesdropper. This was done by showing that principle of indirect decoding introduced by Nair and El Gamal is optimal for this class of channels. This result is in contrast with the non-secrecy case, where the indirect decoding does not provide any gain. 
\end{abstract}

\begin{IEEEkeywords}
broadcast channel, degraded message sets, message cognition, joint secrecy, individual secrecy, capacity regions, more capable channels.
\end{IEEEkeywords}

%
\IEEEpeerreviewmaketitle

\section{Introduction}
The broadcast channel (BC) with degraded message sets was initially introduced by K\"{o}rner and Marton in \cite{DMSKM}. They considered a two-receiver BC, where a common message is transmitted to both receivers and a private message is transmitted to only one of them. They established the capacity region for the general BC by providing a strong converse. The extension of K\"{o}rner and Marton results to the three-receiver BC with two degraded message sets has been considered in \cite{DMSDT1,DMSDT2}, where it has been shown that the straightforward extension of the K\"{o}rner and Marton inner bound is optimal for many special cases. In \cite{DMSNG}, Nair and El Gamal considered a three-receiver BC with degraded message sets, where a common message is sent to all three receivers, while a private message is sent to only one receiver. They showed that the straightforward extension of the K\"{o}rner and Marton inner bound for this scenario is no longer optimal. They presented a new coding scheme known as \textit{indirect decoding} and showed that the resultant inner bound of this technique is strictly greater than the K\"{o}rner and Marton inner bound. However, in \cite{DMSNW}, Nair and Wang showed that if the private message is to be sent to two receivers instead of one, the idea of indirect decoding does not yield any region better than the K\"{o}rner and Marton inner bound. Another scenario for three-receiver BC with degraded message sets was considered in \cite{BCMCT}, where a common message is sent to all three receivers, while two private messages are only sent to two receivers with some message cognition at these receivers. In general, the transmission of degraded message sets over three-receiver BC has captured a lot of attention, yet it has not been  completely solved as many questions remained unanswered beyond the two-receiver case. \newline     
\indent Recent work does not only consider reliable transmission, but it also considers more complex scenarios that involve certain secrecy requirements. In particular, \textit{physical layer security} has attracted a lot of researchers nowadays, see for example \cite{SPLW,Bloch,LiangITS,PLSWC} and references therein. Shannon was the first one to study the problem of secure communication from an information theoretic perspective in \cite{Shann}. He showed that it can be achieved by a secret key shared between the transmitter and the receiver if the entropy of this key is greater than or equal to the entropy of the message to be transmitted. In \cite{Wyn}, Wyner studied the degraded wiretap channel and proved that secure transmission is still achievable over a noisy channel without any secret key. In \cite{Csis}, Csisz\'{a}r and K\"{o}rner extended Wyner's result to the general BC with common and confidential messages. In \cite{SK1,SK2}, the previous two approaches were combined by studying the availability of a shared secret key during secure transmission over a wiretap channel. In \cite{Kang}, Kang and Liu proved that the secrecy capacity for this scenario is achieved by combining the wiretap coding principle along with Shannon's one-time pad idea. Over the years, the integration of confidential and public services over different channels has become very important \cite{RH14}.\newline
\indent Despite the tremendous effort of researchers, the extension of Csisz\'{a}r and K\"{o}rner's work to BC with two or more legitimate receivers has remained an open topic. In \cite{Gamal}, Chia and El Gamal investigated the transmission of one common and one confidential message over a BC with two legitimate receivers and one eavesdropper. They derived a general achievable rate region and established the secrecy capacity if the two legitimate receivers are less noisy than the eavesdropper. They also showed that in some cases the indirect decoding can provide an inner bound that is strictly larger than the direct extension of Csisz\'{a}r and K\"{o}rner's approach. \newline
\indent In this paper we investigate the transmission of degraded message sets with two layers over a three-receiver BC under different secrecy constraints. Our model combines the scenarios in \cite{DMSNW,BCMCT,Gamal} as follows: a common message is transmitted to all three receivers, a confidential common message to the two legitimate receivers and two confidential individual messages to the two legitimate receivers, where each receiver is only interested in one them, while being fully cognizant of the other one. This problem is of high interest and importance because it does not only generalize and combine the previous works in \cite{DMSNW,BCMCT,Gamal}, but it is also of practical relevance since it can be motivated by the concept of two-phase bidirectional relaying in a three-node network \cite{Holger,TPR}.\newline
\indent In the first phase of the bidirectional relaying, node 1 and node 2 transmit their messages to the relay node which decodes them, while keeping the eavesdropper unable to intercept any information about the transmission. This phase corresponds to the multiple access wiretap channel and was investigated in \cite{MAC1,MAC3,MAC2}, where the latter discusses different secrecy criteria. Our work is related to the succeeding broadcast phase, where the relay re-encodes and transmits these messages back to the intended nodes. Since the receiving nodes are cognizant of their own message from the previous phase, they can use it as an additional side information for decoding. First results for the case where this communication scenario must be protected against an additional eavesdropper appeared in \cite{Aydin}, where different achievable rate regions and an outer bound were provided. In our problem, we have an additional feature as the relay transmits another common confidential message to both legitimate receivers and a common message for all three nodes.  \newline
\indent In \cite{Aydin}, the authors claimed to define the secrecy requirement of their model based on a conservative secrecy measure known as \textit{joint secrecy}. This measure assures the secrecy of each confidential message even if the other one is compromised. However, they established an achievable region in \cite[Theorem 1]{Aydin} using secret key approach, where the confidential message of one user is used as a secret key for the other one. One can show that this encoding scheme does not fulfill the joint secrecy constraint. This observation encouraged us to consider another secrecy constraint, in which the legitimate receivers can cooperate together to protect their confidential messages based on some form of mutual trust. This led to the less conservative secrecy measure known as \textit{individual secrecy}. In \cite{Mansour,MansourITW}, we investigated the effect of relaxing the secrecy constraint from joint secrecy to individual secrecy on the capacity region of the BC with receiver side information. On the other hand, different individual secrecy coding techniques has been introduced in an early parallel and independent work in \cite{Ind1} and more recently in \cite{Aydin2,Ind2,Ind3}.\newline
\indent The rest of this paper is organized as follows. In Section \ref{Sec:BCNS}, we introduce the model of three-receiver BC with degraded message sets and full message cognition without any secrecy constraints. We provide a weak converse showing that the straightforward extension of the K\"{o}rner and Marton inner-bound is in fact the capacity region. This result agrees with the one in \cite{DMSNW}, that for this case indirect decoding can not outperform the K\"{o}rner and Marton inner bound. In Section \ref{Sec:BCSC}, we introduce secrecy constraints to our model and discuss the differences between the joint and individual secrecy criteria. In Section \ref{Sec:BCJS}, we provide an achievable rate region for the joint secrecy criterion. We then establish the joint secrecy capacity region if only one of the legitimate receivers is more capable than the eavesdropper using the principle of indirect decoding. In Section \ref{Sec:BCIS}, we provide an achievable rate region for the individual secrecy criterion. We then establish the individual secrecy capacity region if the two legitimate receivers are more capable than the eavesdropper. 
\subsection*{Notation}
In this paper, random variables are denoted by capital letters and their realizations by the corresponding lower case letters, while calligraphic letters are used to denote sets. $\mathrm{X}^n$ denotes the sequence of variables $(\mathrm{X}_1,\ldots ,\mathrm{X}_n)$, where $\mathrm{X}_i$ is the $i^{th}$ variable in the sequence. Additionally, we use $\tilde{\mathrm{X}}^i$ to denote the sequence $(\mathrm{X}_i,\ldots,\mathrm{X}_n$). A probability distribution for the random variable X is denoted by $Q(x)$. $\mathrm{U - V - X}$ denotes a Markov chain of random variable U, V and X in this order, while $\mathrm{(U - V, K) - X - Y}$ implies that $\mathrm{U - V - X - Y}$ and $\mathrm{K - X - Y}$ are Markov chains. $\mathbb{R}_+$ is used to denote the set of nonnegative real numbers. $\mathbb {H}(\cdot)$ and $\mathbb{I}(\cdot;\cdot)$ are the traditional entropy and mutual information. The probability of an event is given by $\mathbb{P}[\cdot]$, while $\mathbb{E}[\cdot]$ is used to represent the expectation. Moreover, $\llbracket a,b \rrbracket$ is used to represent the set of natural numbers between $a$ and $b$.
\section{BC with Degraded Message Sets and Message Cognition}
\label{Sec:BCNS}
In this section, we investigate the three-receiver BC with degraded message sets and message cognition without any secrecy constraints. First, we introduce our model, then establish the capacity region for the general three-receiver BC with two degraded message sets.
\subsection{System Model and Channel Comparison}
\begin{figure}[h]
\centering
\includegraphics[scale=0.68]{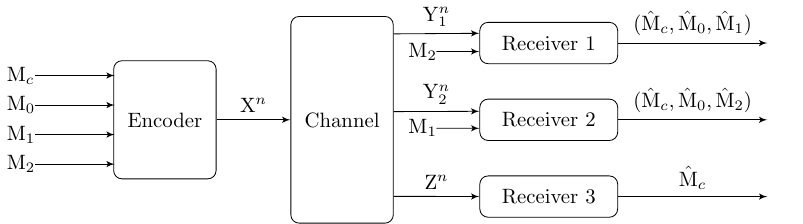}
\caption{Three-receiver broadcast channel with degraded message sets and message cognition.}
\label{Fig:BCNS}
\end{figure}

Let $\mathcal{X}$, $\mathcal{Y}_1$, $\mathcal{Y}_2$ and $\mathcal{Z}$ be finite input and output sets. Then for input and output sequences $x^n \in \mathcal{X}^n$, $y_1^n \in \mathcal{Y}_1^n$, $y_2^n \in \mathcal{Y}_2^n$ and $z^n \in \mathcal{Z}^n$ of length $n$, the discrete memoryless BC is given by
\begin{equation*}
Q^n(y_1^n , y_2^n , z^n \arrowvert x^n) = \prod_{k=1}^n Q({y_1}_k, {y_2}_k, z_k \arrowvert x_k),
\end{equation*} 
where $x^n$ represents the transmitted sequence, $y_1^n$, $y_2^n$ and $z^n$ represent the received sequence at the three receivers. Before we discuss our model in details, we need to introduce two important classes of BCs, that we will address a lot in our investigation. 
\begin{mydef}
In a discrete memoryless BC $Q(y,z\arrowvert x)$, $\mathrm{Y}$ is said to be less noisy than $\mathrm{Z}$, also written as $\mathrm{Y \succeq Z}$, if for every random variable $\mathrm{V}$ such that $\mathrm{V - X - (Y,Z)}$ forms a Markov chain, we have
\begin{equation}
\mathrm{\mathbb{I}(V;Y}) \geq \mathrm{\mathbb{I}(V;Z)}. 
\label{Equ:LessN}
\end{equation}
On the other hand, $\mathrm{Y}$ is said to be more capable than $\mathrm{Z}$, if for every input distribution on $\mathrm{X}$, we have
\begin{equation}
\mathrm{\mathbb{I}(X;Y}) \geq \mathrm{\mathbb{I}(X;Z)}. 
\label{Equ:MoreCap}
\end{equation}
\end{mydef}
The class of more capable channels is strictly wider than the less noisy one. It can be shown that any less noisy channel is a more capable one. Further, it was shown that the class of less noisy channels contains the physically and stochastically degraded channels \cite{Gamal2}. \newline
\indent  We consider the standard model with a block code of arbitrary but fixed length $n$. We consider four different messages sets. The first set contains the common messages for all three receivers and is denoted by $\mathcal{M}_c = \llbracket 1,2^{nR_c} \rrbracket$. The second set is denoted by $\mathcal{M}_0 = \llbracket 1,2^{nR_0} \rrbracket$ and contains the private common messages for Receivers $1$ and $2$. While the last two sets contain the individual private messages $\mathcal{M}_1 = \llbracket 1,2^{nR_1} \rrbracket$ and $\mathcal{M}_2 = \llbracket 1,2^{nR_2} \rrbracket$. Further, we assume full message cognition at $\mathrm{Y}_1$ and $\mathrm{Y}_2$\footnote{From this point, we will refer to different receivers by their respective channel outputs interchangeably.}, such that $\mathrm{Y}_1$ is cognizant of the entire message $\mathrm{M}_2$ and $\mathrm{Y}_2$ of the entire message $\mathrm{M}_1$ as shown in Fig. \ref{Fig:BCNS}.
\begin{Remark}
If we let $\mathcal{M}_1= \mathcal{M}_2 = \emptyset$, our model reduces to the three-receiver BC with two degraded message sets studied in \cite{DMSNG,DMSNW}.
\end{Remark}
\begin{mydef}
A $(2^{nR_c},2^{nR_0},2^{nR_1},2^{nR_2},n)$ code $\mathcal{C}_n$ for the BC with degraded message sets and message cognition consists of: four independent message sets $\mathcal{M}_c$, $\mathcal{M}_0$, $\mathcal{M}_1$ and $\mathcal{M}_2$; an encoding function at the transmitter 
\begin{equation*}
E :\mathcal{M}_c \times \mathcal{M}_0 \times \mathcal{M}_1 \times \mathcal{M}_2 \rightarrow \mathcal{X}^n
\end{equation*} 
which maps a message quadruple $(m_c,m_0,m_1,m_2) \in \mathcal{M}_c \times \mathcal{M}_0 \times \mathcal{M}_1 \times \mathcal{M}_2$ to a codeword $x^n(m_c,m_0,m_1,m_2)$; and three decoders, one at each receiver
\begin{align*}
\varphi_1 &: \mathcal{Y}_1^n \times \mathcal{M}_2 \rightarrow \mathcal{M}_c \times \mathcal{M}_0 \times \mathcal{M}_1  \cup \{?\} \\ 
\varphi_2 &: \mathcal{Y}_2^n \times \mathcal{M}_1 \rightarrow \mathcal{M}_c \times \mathcal{M}_0 \times \mathcal{M}_2  \cup \{?\} \\
\varphi_3 &: \mathcal{Z}^n \rightarrow \mathcal{M}_c \cup \{?\}  
\end{align*} 
that maps each channel observation at the respective receiver and the cognizant message to the corresponding intended messages or an error message $\{?\}$.   
\end{mydef}
We assume that the messages $\mathrm{M}_c$, $\mathrm{M}_0$, $\mathrm{M}_1$ and $\mathrm{M}_2$ are independent and chosen uniformly at random. The reliability performance of $\mathcal{C}_n$ is measured in terms of its average probability of error
\begin{align}
P_e(\mathcal{C}_n) &\triangleq \mathbb{P} \Big[ \mathrm{ (\hat{M}_c, \hat{M}_0, \hat{M}_1) \neq (M_c, M_0, M_1) \text{ or }} \nonumber\\
&\mathrm{(\tilde{M}_c, \tilde{M}_0,\tilde{M}_2) \neq (M_c, M_0, M_2) \text{ or } \check{M}_c \neq M_c} \Big],
\label{Equ:ErrorProb}
\end{align}
where $\mathrm{(\hat{M}_c, \hat{M}_0, \hat{M}_1)}$, $\mathrm{(\tilde{M}_c, \tilde{M}_0, \tilde{M}_2)}$ and $\mathrm{\check{M}_c}$ are the estimated messages at $\mathrm{Y}_1$, $\mathrm{Y}_2$ and $\mathrm{Z}$ respectively. 
\begin{mydef}
A rate quadruple $(R_c,R_0,R_1,R_2) \in \mathbb{R}_+^4$ is achievable for the BC with degraded message sets and message cognition, if there exists a sequence of $(2^{nR_c},2^{nR_0},2^{nR_1}, 2^{nR_2}, n)$ codes $\mathcal{C}_n$ and a sequence $\epsilon_n$, such that for $n$ is large enough, the following holds         
\begin{equation}
P_e (\mathcal{C}_n) \leq \epsilon_n \hspace{1cm} \text{and} \hspace{1cm} \lim_{n \rightarrow \infty} \epsilon_n = 0. 
\label{Equ:NSAchCond}
\end{equation}
\end{mydef}  

\subsection{Capacity Region}
\begin{mytheo}
The capacity region of the three-receiver BC with degraded message sets and message cognition is the set of all rate quadruples $(R_c,R_0,R_1,R_2) \in \mathbb{R}_+^4$ that satisfy
\begin{align}
R_c &\leq \mathrm{\mathbb{I} (U;Z)} \nonumber \\
R_0 + R_1 &\leq \mathrm{\mathbb{I}(X;Y_1\arrowvert U)} \nonumber \\
R_0 + R_2 &\leq \mathrm{\mathbb{I}(X;Y_2\arrowvert U)} \nonumber \\
R_c + R_0 + R_1 &\leq \mathrm{\mathbb{I}(X;Y_1)} \nonumber \\
R_c + R_0 + R_2 &\leq \mathrm{\mathbb{I}(X;Y_2)}
\label{Equ:NSCap} 
\end{align}
for some $\mathrm{(U,X)}$, such that $\mathrm{U - X - (Y_1, Y_2, Z)}$ forms a Markov chain. Further it suffices to have $\arrowvert \mathcal{U} \arrowvert \leq \arrowvert \mathcal{X} \arrowvert + 2$.
\label{Ther:BCNS}
\end{mytheo}

\begin{IEEEproof}
The achievability follows directly from the straightforward extension of the K\"{o}rner and Marton inner bound in \cite{DMSKM} to the three-receiver BC with degraded message sets and message cognition as in \cite{DMSNW,BCMCT}. Superposition encoding is used as follows: $m_c$ is encoded in the cloud centers codewords $\mathrm{U}^n$, while $(m_0,m_1,m_2)$ are superimposed in the satellite codewords $\mathrm{X}^n$. Joint typicality decoders are then used at each receiver leading to the bounds in \eqref{Equ:NSCap}. \newline
\indent For the converse, we start by establishing the reliability upper bounds for any achievable rates. Based on Fano's inequality, the expression of the average error probability in \eqref{Equ:ErrorProb} and the reliability constraint given by \eqref{Equ:NSAchCond}, we have
\begin{align}
\mathbb{H}(\mathrm{M}_c \arrowvert \mathrm{Z}^n), \mathbb{H}( \mathrm{M}_c \arrowvert \mathrm{Y}_1^n \mathrm{M}_2), \mathbb{H}(\mathrm{M}_c \arrowvert \mathrm{Y}_2^n \mathrm{M}_1) &\leq n\gamma_c (\epsilon_n) \label{Equ:RelComCod} \\
\mathbb{H}(\mathrm{M_0 M_1} \arrowvert \mathrm{Y}_1^n \mathrm{M}_2 \mathrm{M}_c) &\leq n\gamma_1 (\epsilon_n) \label{Equ:RelFirCond} \\
\mathbb{H}(\mathrm{M_0 M_2} \arrowvert \mathrm{Y}_2^n \mathrm{M}_1 \mathrm{M}_c) &\leq n\gamma_2 (\epsilon_n) \label{Equ:RelSecCond}
\end{align}
where $\gamma_c(\epsilon_n)= 1/n + \epsilon_n R_c$, $\gamma_1(\epsilon_n) = 1/n + \epsilon_n(R_0 + R_1)$ and $\gamma_2(\epsilon_n) = 1/n + \epsilon_n(R_0 + R_2)$. \newline
\indent Next, we let $\mathrm{U}_i \triangleq (\mathrm{M}_c,\tilde {\mathrm{Z}}^{i+1})$, $\mathrm{K}_i^1 \triangleq \mathrm{Y}_1^{i-1}$, $\mathrm{K}_i^2 \triangleq \mathrm{Y}_2^{i-1}$, $\mathrm{M \triangleq (M_0, M_1, M_2)}$, $\mathrm{V}_i \triangleq (\mathrm{M},\mathrm {U}_i)$, $\mathrm{V}_i^1 \triangleq (\mathrm{V}_i, \mathrm{K}_i^1)$ and $\mathrm{V}_i^2 \triangleq (\mathrm{V}_i, \mathrm{K}_i^2)$. We then start by considering the common rate $R_c$. Using Eq. \eqref{Equ:RelComCod}, we have
\begin{align}
R_c &\leq \frac{1}{n} \Big[ \mathbb{H}(\mathrm{M}_c) - \mathbb{H} (\mathrm{M}_c \arrowvert \mathrm{Z}^n) \Big] + \gamma_c(\epsilon_n) \nonumber\\
&= \frac{1}{n} \mathbb{I} (\mathrm{M}_c;\mathrm{Z}^n) + \gamma_c(\epsilon_n) \nonumber \\
&= \frac{1}{n} \sum_{i=1}^n \mathbb{I}(\mathrm{M}_c;\mathrm{Z}_i \arrowvert \tilde{\mathrm{Z}}^{i+1}) + \gamma_c(\epsilon_n) \nonumber \\
&\leq \frac{1}{n} \sum_{i=1}^n \mathbb{I}(\mathrm{M}_c \tilde{\mathrm{Z}}^{i+1} ;\mathrm{Z}_i) + \gamma_c(\epsilon_n) \nonumber \\
&= \frac{1}{n} \sum_{i=1}^n \mathbb{I} (\mathrm{U}_i;\mathrm{Z}_i) + \gamma_c(\epsilon_n). 
\label{Equ:ConvNS1_1}
\end{align}
Next, we consider the sum of the private rates $(R_0 + R_1)$ which are intended for receiver $\mathrm{Y}_1$. We have
\begin{align}
R_0 &+ R_1 \overset{(a)}{\leq} \frac{1}{n} \mathrm{\mathbb{I}(M_0 M_1;Y}_1^n \arrowvert \mathrm{M}_2 \mathrm{M}_c) + \gamma_1(\epsilon_n) \nonumber\\
&\overset{(b)}{\leq} \frac{1}{n} \Big[\mathrm{\mathbb{I}(M;Y}_1^n \arrowvert \mathrm{M}_c) + \mathrm{\mathbb{I}(M;Z}^n \arrowvert \mathrm{M}_c) - \mathrm{\mathbb{I}(M;Z}^n \arrowvert \mathrm{M}_c) \Big]\nonumber\\
&\qquad + \gamma_1(\epsilon_n) \nonumber\\
&= \frac{1}{n} \sum_{i=1}^n \Big[ \mathrm{\mathbb{I}(M;Y}_{1i} \arrowvert \mathrm{M}_c \mathrm{Y}_1^{i-1}) + \mathrm{\mathbb{I}(M;Z}_i \arrowvert \mathrm{M}_c \tilde{\mathrm{Z}} ^{i+1}) \nonumber\\
&\qquad  - \mathrm{\mathbb{I}(M;Z}_i \arrowvert \mathrm{M}_c \tilde{\mathrm{Z}}^{i+1})\Big] + \gamma_1 (\epsilon_n) \nonumber\\
&\overset{(c)}{=} \frac{1}{n} \sum_{i=1}^n \Big[ \mathrm{\mathbb{I}(M; Y}_{1i} \arrowvert \mathrm{M}_c \mathrm{Y}_1^{i-1} \tilde{\mathrm{Z}}^{i+1}) + \mathrm{\mathbb{I}(M;Z}_i \arrowvert \mathrm{M}_c \tilde{\mathrm{Z}}^{i+1})\nonumber\\
&\qquad - \mathrm{\mathbb{I}(M;Z}_i \arrowvert \mathrm{M}_c \mathrm{Y}_1^{i-1} \tilde{\mathrm{Z}}^{i+1}) \Big] + \gamma_1 (\epsilon_n) \nonumber\\
&= \frac{1}{n} \sum_{i=1}^n \Big[\mathbb{I}(\mathrm{V}_i^1; \mathrm{Y}_{1i} \arrowvert \mathrm{U}_i \mathrm{K}_i^1) - \mathbb{I}(\mathrm{V}_i^1;\mathrm{Z}_i \arrowvert \mathrm{U}_i \mathrm{K}_i^1) \nonumber\\
&\qquad + \mathbb{I}(\mathrm{V}_i;\mathrm{Z}_i \arrowvert \mathrm{U}_i) \Big] + \gamma_1(\epsilon_n), 
\label{Equ:ConvNS1_2}
\end{align}
where $(a)$ follows from \eqref{Equ:RelFirCond}; $(b)$ follows as $\mathrm{\mathbb{I}(M;Y}_1^n \arrowvert \mathrm{M}_c) \geq \mathrm{ \mathbb{I}(M_0 M_1;Y}_1^n \arrowvert \mathrm{M}_2 \mathrm{M}_c)$ and $(c)$ follows by the Csisz\'{a}r sum identity \cite[Lemma 7]{Csis}. If we use Eq. \eqref{Equ:RelSecCond} and follow the exact same steps, we can derive a similar bound for the sum of the private rates $(R_0 + R_2)$ intended for receiver $\mathrm{Y}_2$ as follows:
\begin{align}
R_0 + R_2 &\leq \frac{1}{n} \sum_{i=1}^n \Big[\mathbb{I}(\mathrm{V}_i^2; \mathrm{Y}_{2i} \arrowvert \mathrm{U}_i \mathrm{K}_i^2) - \mathbb{I}( \mathrm{V}_i^2;\mathrm{Z}_i \arrowvert \mathrm{U}_i \mathrm{K}_i^2) \nonumber\\
&\qquad + \mathbb{I}(\mathrm{V}_i;\mathrm{Z}_i \arrowvert \mathrm{U}_i)\Big] + \gamma_2(\epsilon_n). 
\label{Equ:ConvNS1_3}
\end{align}
Now using \eqref{Equ:ConvNS1_1}, \eqref{Equ:ConvNS1_2} and \eqref{Equ:ConvNS1_3} followed by introducing a time sharing random variable $\mathrm{T}$ independent of all others and uniformly distributed over $\llbracket 1; n \rrbracket$, and let $\mathrm{U = (U_T,T)}$, $\mathrm{K^1 = K_T^1}$, $\mathrm{K^2 = K_T^2}$, $\mathrm{V = V_T}$, $\mathrm{V^1 = V_T^1}$, $\mathrm{V^2 = V_T^2}$, $\mathrm{Y_1 = Y_{1T}}$, $\mathrm{Y_2 = Y_{2T}}$ and $\mathrm{Z = Z_T}$, then take the limit as $n \rightarrow \infty$ such that, $\gamma_c(\epsilon_n)$, $\gamma_1(\epsilon_n)$ and $\gamma_2(\epsilon_n) \rightarrow 0$, we reach the following
\begin{subequations} \label{Equ:ConvNS1_4}
\begin{align}
R_c &\leq \mathrm{\mathbb{I} (U;Z)} \label{Equ:ConvNS1_4_1} \\
R_0 + R_1 &\leq \mathrm{\mathbb{I}(V^1;Y_{1} \arrowvert U K^1) - \mathbb{I}(V^1;Z \arrowvert U K^1) + \mathbb{I}(V;Z \arrowvert U)} \label{Equ:ConvNS1_4_2} \\
R_0 + R_2 &\leq \mathrm{\mathbb{I}(V^2;Y_{2} \arrowvert U K^2) - \mathbb{I}(V^2;Z \arrowvert U K^2) + \mathbb{I}(V;Z \arrowvert U)} \label{Equ:ConvNS1_4_3},
\end{align}
\end{subequations}
where $\sq{\mathrm{(U - V, K^1) - V^1 - X - (Y_{1},Y_{2},Z_)}}$ and $\sq{\mathrm{(U - V,K^2)- V^2}}$ $\sq{\mathrm{- X - (Y_{1},Y_{2},Z)}}$ form Markov chains. Since the conditional mutual information is the expectation of the unconditional one, Eq. \eqref{Equ:ConvNS1_4_2} can be further upper-bounded as follows: 
\begin{align}
\sq{R_0 + R_1} &\leq \sq{\mathbb{E}_\mathrm{K^1} \Big[\mathrm{\mathbb{I}(V^1;Y_{1} \arrowvert U K^1)} - \mathrm{\mathbb{I}(V^1;Z \arrowvert U K^1)\Big]}}+ \mathrm{\mathbb{I}(V; Z \arrowvert U)} \nonumber\\
&\overset{(a)} {\leq} \sq{\mathrm{\mathbb{I}(V^1;Y_{1} \arrowvert U, \space K^1}=k^{1*}) - \mathrm{\mathbb{I} (V^1;Z \arrowvert U, \space K^1} = k^{1*})} \nonumber\\
&\qquad + \mathrm{\mathbb{I}(V;Z \arrowvert U)} \nonumber\\
&\overset{(b)} {=} \mathrm{\mathbb{I}(V^{1*};Y_{1} \arrowvert U)} - \mathrm{\mathbb{I} (V^{1*};Z \arrowvert U)} + \mathrm{\mathbb{I}(V;Z \arrowvert U)} \label{Equ:ConvNS1_5}
\end{align} 
where $(a)$ follows as $k^{1*}$ is the value of $\mathrm{K^1}$ that maximizes the difference $\sq{\mathrm{\mathbb{I}(V^1;Y_{1} \arrowvert U, \space K^1}=k^1) - \mathrm{ \mathbb{I}(V^1;Z \arrowvert U, \space K^1} = k^1)}$; while $(b)$ follows because $\mathrm {V^{1*}}$ is distributed according to the following probability distribution $\sq{Q(v^1 \arrowvert u, k^1= k^{1*})}$ \cite[Corollary 2.3]{LiangITS}. This implies that the right hand side of Eq. \eqref {Equ:ConvNS1_4_2} is maximized by setting $\mathrm{K^1}= k^{1*}$. Using this result, we can upper-bound Eq. \eqref {Equ:ConvNS1_4_2} as follows:  
\begin{align}
R_0+R_1 &\leq \mathrm{\mathbb{I}(V^1;Y_{1} \arrowvert U K^1) - \mathbb{I}(V^1;Z \arrowvert U K^1) + \mathbb{I}(V;Z \arrowvert U)} \nonumber\\
&\overset{(a)}{=}\mathrm{\mathbb{I}(V^1;Y_{1} \arrowvert U K^1) + \mathbb{I}(K^1;Z \arrowvert U) - \mathbb{I}(K^1;Z \arrowvert V)} \nonumber\\ 
&\overset{(b)}{\leq} \mathrm{\mathbb{I}(V^1;Y_{1} \arrowvert U,\space K^1}=k^{1*}) + \mathbb{I} (\mathrm{K^1} = k^{1*}; \mathrm{Z \arrowvert U)} \nonumber\\
&\qquad - \mathbb{I}(\mathrm{K^1} = k^{1*};\mathrm{Z \arrowvert V)} \nonumber\\
&\overset{(c)} {=} \mathrm{\mathbb{I}(V^{1*};Y_{1} \arrowvert U)} \nonumber\\
&\overset{(d)} {\leq} \mathrm{\mathbb{I}(X;Y_{1} \arrowvert U)}, \label{Equ:ConvNS1_7}
\end{align}
where $(a)$ follows by the mutual information chain rule; $(b)$ follows because setting  $\mathrm{K^1}= k^{1*}$ maximizes the right hand side of Eq. \eqref {Equ:ConvNS1_4_2}; $(c)$ follows because $\mathbb{I}(k^{1*}; \mathrm{Z \arrowvert U)}$ and $\mathbb{I} (k^{1*};\mathrm{Z \arrowvert V)}$ vanish for a fixed realization of $\mathrm{K^1} = k^{1*}$; while $(d)$ follows from the data processing inequality and the fact that $\mathrm{U - V^{1*} - X -}\mathrm{(Y_{1},Y_{2},Z)}$ forms a Markov chain, which implies that $\mathrm{ \mathbb{I}(V^{1*};Y_{1} \arrowvert U) \leq \mathbb{I}(X;Y_{1} \arrowvert U)}$. \newline
\indent Now, If we apply the same steps and ideas to  Eq. \eqref{Equ:ConvNS1_4_3}, we can derive the following bound: 
\begin{align}
R_0 + R_2 &\leq \mathrm{\mathbb{I}(V^{2*};Y_{2} \arrowvert U)} \nonumber\\
&\leq \mathrm{\mathbb{I}(X;Y_{2} \arrowvert U)}, \label{Equ:ConvNS1_8}
\end{align} 
where $\mathrm{U - V^{2*} - X - (Y_{1},Y_{2},Z)}$ forms a Markov chain and $\mathrm {V^{2*}}$ is distributed as $Q(v^2 \arrowvert u,k^2= k^{2*})$ such that, $k^{2*}$ is the value of $\mathrm{K}^2$ that maximizes the difference $\mathrm{\mathbb{I}(V^2;Y_{2} \arrowvert U, K^2}=k^2) - \mathrm{ \mathbb{I}(V^2;Z \arrowvert U, K^2} = k^2)$. At this point we need to illustrate an important fact. One might argue that getting rid of the two conditional random variables $\mathrm {K^1}$ and $\mathrm{K^2}$ as we did, can not be done simultaneously because $\mathrm{K^1}$ and $\mathrm{K^2}$ might be dependent, such that the maximizing values $k^{1*}$ and $k^{2*}$ can not occur concurrently. However, this argument does not affect our converse because it only implies that the derived upper bounds might not be as tight as the original ones. To finalize our converse, we need to highlight the standard upper bounds for reliable transmission
\begin{align}
R_c + R_0 + R_1 &\leq \mathrm{\mathbb{I}(X;Y_1)} \nonumber\\
R_c + R_0 + R_2 &\leq \mathrm{\mathbb{I}(X;Y_2)}. \label{Equ:ConvNS1_9}
\end{align} 
\indent Now, if we combine \eqref{Equ:ConvNS1_4_1}, along with \eqref{Equ:ConvNS1_7}, \eqref{Equ:ConvNS1_8} and \eqref{Equ:ConvNS1_9}, such that $\mathrm{U - X - (Y_1,Y_2,Z)}$ forms a Markov chain, we reach the same region given by \eqref{Equ:NSCap}. In order to complete our converse, we need to point out that the cardinality argument $\arrowvert \mathcal{U} \arrowvert \leq \arrowvert \mathcal{X} \arrowvert + 2$ follows from the Fenchel-Bunt strengthening of the usual Carath\'{e}odory's theorem \cite[Appendix C]{Gamal2}.
\end{IEEEproof}
\begin{Remark}
It is important to note that, Theorem \ref{Ther:BCNS} implies that the inner bound established in \cite{DMSNW} is in fact the capacity region. 
\end{Remark}
\section{Secrecy in BC with Degraded Message Sets and Message Cognition}
\label{Sec:BCSC}
In this section, we will investigate the three-receiver BC with degraded message sets and message cognition under two different secrecy constraints: joint secrecy and individual secrecy. We compare these two criteria by investigating their capacity regions for some special cases and show that the individual secrecy provides a larger secrecy capacity compared to the joint one.

\subsection{Secrecy Model and Criteria}
We start by modifying the model introduced in the previous section, such that the private messages $\mathrm{M_0, M_1}$ and $\mathrm{M}_2$ are now confidential messages that need to be kept secret from the eavesdropper as shown in Figure \ref{Fig:BCWS}. Our new code is defined as follows: 
\begin{figure} [h]
\centering
\includegraphics[scale=0.65]{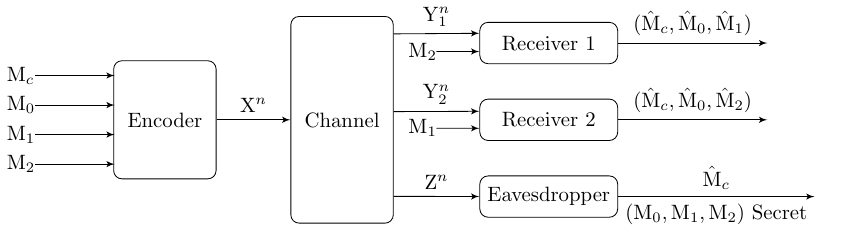}
\caption{Wiretap broadcast channel with degraded message sets and message cognition}
\label{Fig:BCWS}
\end{figure}
\begin{mydef}
A $(2^{nR_c},2^{nR_0},2^{nR_1},2^{nR_2},n)$ code $\mathcal{C}^s_n$ for the wiretap BC with degraded message sets and message cognition consists of: four independent message sets $\mathcal{M}_c$, $\mathcal{M}_0$, $\mathcal{M}_1$ and $\mathcal{M}_2$; a source of local randomness at the encoder $\mathcal{R}$ which is distributed according to $Q(r)$; an encoding function at the relay node 
\begin{equation*}
E :\mathcal{M}_c \times \mathcal{M}_0 \times \mathcal{M}_1 \times \mathcal{M}_2 \times \mathcal{R} \rightarrow \mathcal{X}^n
\end{equation*} 
which maps a common message $m_c \in \mathcal{M}_c$, a confidential message triple $(m_0,m_1 , m_2) \in \mathcal{M}_0 \times \mathcal{M}_1 \times \mathcal{M}_2$ and a realization of the local randomness $r \in \mathcal{R}$ to a codeword $x^n(m_c,m_0,m_1,m_2,r)$, and three decoders, one for each node
\begin{align*}
\varphi_1 &: \mathcal{Y}_1^n \times \mathcal{M}_2 \rightarrow \mathcal{M}_c \times \mathcal{M}_0 \times \mathcal{M}_1  \cup \{?\} \\ 
\varphi_2 &: \mathcal{Y}_2^n \times \mathcal{M}_1 \rightarrow \mathcal{M}_c \times \mathcal{M}_0 \times \mathcal{M}_2  \cup \{?\} \\
\varphi_3 &: \mathcal{Z}^n \rightarrow \mathcal{M}_c \cup \{?\}  
\end{align*} 
that maps each channel observation at the respective node and the cognizant message to the corresponding required messages or an error message $\{?\}$.   
\end{mydef}
We assume that the messages $\mathrm{M}_c$, $\mathrm{M}_0$, $\mathrm{M}_1$ and $\mathrm{M}_2$ are chosen uniformly at random and use the average error probability in \eqref{Equ:ErrorProb} to measure the reliability performance of the code $\mathcal{C}^s_n$. On the other hand, the secrecy performance of $\mathcal{C}^s_n$ is measured with respect to two different criteria. These two criteria identify the level of ignorance of the eavesdropper\footnote{Although the third receiver $(\mathrm{Z})$ is part of our model and not an external user, we will refer to it in the rest of the paper as an eavesdropper.} about the confidential messages $\mathrm{M}_0 $, $\mathrm{M}_1$ and $\mathrm{M}_2$ as follows: \vspace{0.2cm} \newline
\indent \underline{\textbf{1. Joint Secrecy:}} This criterion requires the leakage of the confidential messages of one user to the eavesdropper given the individual message of the other user to be small. For our model, this requirement can be expressed as follows:
\begin{gather}
\mathbb{I} \mathrm{(M_0 M_1};\mathrm{Z}^n \arrowvert \mathrm{M_2}) \leq \tau_{1n} \hspace{0.5cm} \text{and} \hspace{0.5cm}  \mathbb{I} \mathrm{(M_0 M_2};\mathrm{Z}^n \arrowvert \mathrm{M_1}) \leq \tau_{2n},\nonumber \\
\text{where } \lim_{n \rightarrow \infty} \tau_{1n},\tau_{2n} = 0.
\label{Equ:JoiCond}  
\end{gather}
This criterion guarantees that the rate of information leaked to the eavesdropper from one user is small even if the other individual transmitted message is compromised. Thus, in this scenario the legitimate receivers do not have to trust each other. In some literature, the joint secrecy criterion is defined such that, the mutual leakage of all confidential messages to the eavesdropper is small as follows:
\begin{equation}
\mathbb{I} \mathrm{(M_0 M_1 M_2 ;Z}^n) \leq \tau_n \hspace{0.5cm} \text{and} \hspace{0.5cm}  \lim_{n \rightarrow \infty} \tau_{n} = 0.\label{Equ:MJoiSec}
\end{equation}
One can easily show that the definition in \eqref{Equ:JoiCond} is equivalent to the one in \eqref{Equ:MJoiSec} for some $\tau_n$ as follows:
\begin{align*}
\mathbb{I} \mathrm{(M_0 M_1 M_2 ;Z}^n) &= \mathbb{I} \mathrm{(M_0 M_1};\mathrm{Z}^n \arrowvert \mathrm{M_2}) + \mathbb{I} \mathrm{(M_2};\mathrm{Z}^n) \\ 
&\overset{(a)}{\leq} \mathbb{I} \mathrm{(M_0 M_1};\mathrm{Z}^n \arrowvert \mathrm{M_2}) + \mathbb{I} \mathrm{(M_2};\mathrm{Z}^n \arrowvert \mathrm{M_1}) \\
&\overset{(b)}{\leq} \mathbb{I} \mathrm{(M_0 M_1};\mathrm{Z}^n \arrowvert \mathrm{M_2}) + \mathbb{I} \mathrm{(M_0 M_2};\mathrm{Z}^n \arrowvert \mathrm{M_1}) \\ 
&\leq \tau_{1n} + \tau_{2n} \leq \tau_n, 
\end{align*}
where $(a)$ follows because $\mathrm{M}_1$ and $\mathrm{M}_2$ are independent which implies that $ \mathbb{I} \mathrm{(M_2};\mathrm{Z}^n) \leq  \mathbb{I} \mathrm{(M_2};\mathrm{Z}^n \arrowvert \mathrm{M_1})$; while $(b)$ follows because $ \mathbb{I} \mathrm{(M_2};\mathrm{Z}^n \arrowvert \mathrm{M_1}) \leq \mathbb{I} \mathrm{(M_0 M_2};\mathrm{Z}^n \arrowvert \mathrm{M_1})$. On the other hand, if Eq. \eqref{Equ:MJoiSec} holds, it follows directly that $\mathbb{I} \mathrm{(M_0 M_1}; \mathrm{Z}^n \arrowvert \mathrm{M_2}) \leq \tau_n$ and $\mathbb{I} \mathrm{(M_0 M_2};\mathrm{Z}^n \arrowvert \mathrm{M_1}) \leq \tau_n$. However, we prefer the definition in \eqref{Equ:JoiCond}, because it provides a better understanding to the relation between the legitimate receivers and allows us to interpret the immunity of the joint secrecy against compromised receivers. \vspace{0.2cm} \newline
\indent \underline{\textbf{2. Individual Secrecy:}} This criterion requires the leakage of the confidential messages of each user to the eavesdropper to be small without conditioning on the confidential messages of the others users. This requirement can be formulated as follows: 
\begin{equation}
\sq{\mathbb{I} \mathrm{(M_0 M_1};\mathrm{Z}^n) \leq \tau_{1n} \hspace{0.5cm} \text{and} \hspace{0.5cm}  \mathbb{I} \mathrm{(M_0 M_2};\mathrm{Z}^n ) \leq \tau_{2n},} 
\label{Equ:IndCond}  
\end{equation}
where $\tau_{1n}$ and $\tau_{2n}$ are defined as before. Differently from the conservative constraint in \eqref{Equ:JoiCond}, where different users do not trust each other, this secrecy measure allows the legitimate receivers to cooperate in protecting their messages against eavesdropping. In some literatures the individual secrecy criterion requires the sum of the leakages of each confidential message to the eavesdropper to be small as:
\begin{equation}
\mathbb{I} \mathrm{(M_0 ;Z}^n) + \mathbb{I} \mathrm{(M_1 ;Z}^n) + \mathbb{I} \mathrm{(M_2 ;Z}^n) \leq \tau_n.
\label{Equ:IndSec}  
\end{equation}
However, this definition is only equivalent to the one in \eqref{Equ:IndCond} if $\mathrm{M}_0 = \emptyset$, but in general they are not the same. In fact, the constraint in \eqref{Equ:IndCond} is stronger than this one. This is because Eq. \eqref{Equ:IndCond} directly implies Eq. \eqref {Equ:IndSec}, while the opposite is not correct. The difference between these two definitions is in the interpretation of the word individual. In \eqref{Equ:IndCond}, individuality means different transmission flows, while in \eqref{Equ:IndSec} it means different confidential messages. In this paper, we will use the individual secrecy constraint given in \eqref{Equ:IndCond} because it implies the other constraint in \eqref{Equ:IndSec} and we think it is more convenient and meaningful.

\begin{mydef}
A rate quadruple $(R_c,R_0,R_1,R_2) \in \mathbb{R}_+^4$ is achievable for the wiretap BC with degraded message sets and message cognition, if there exist a sequence of $(2^{nR_c},2^{nR_0}, 2^{nR_1},2^{nR_2}, n)$ codes $\mathcal{C}^s_n$ and three sequences $\epsilon_n,\tau_{1n}, \tau_{2n}$, where $n$ is large enough, such that         
\begin{equation}
P_e (\mathcal{C}_n) \leq \epsilon_n, \hspace{1.5cm} \lim_{n \rightarrow \infty} \epsilon_n,\tau_{1n},\tau_{2n}= 0.
\label{Equ:AchCond}
\end{equation}
and depending on the selected secrecy criterion, the conditions in \eqref{Equ:JoiCond} or \eqref{Equ:IndCond} are fulfilled.
\end{mydef} 
\begin{Remark}
It is worth mentioning that the previous definition and the requirements of the joint and individual secrecy criteria use the notation of strong secrecy \cite{StrongSec1,StrongSec}, where the intuition is to have the total amount of information leaked to the eavesdropper to be small.
\end{Remark} 
\begin{Remark} It is important to note how our model generalize different works on the wiretap BC with more than one legitimate receiver as follows: \newline
\indent $\bullet$ If we let $\mathcal{M}_1=\mathcal{M}_2 = \emptyset$, our model reduces to the three receivers BC with common and confidential messages investigated in \cite{Gamal}. \vspace{0.2cm} \newline
\indent $\bullet$ If we let $\mathcal{M}_c=\mathcal{M}_0 = \emptyset$, our model reduces to the wiretap BC with receiver side information. This channel was investigated under the joint secrecy constraint in \cite{Aydin,Mansour} and under the individual secrecy constraint in \cite{Ind1,Mansour}.
\end{Remark}
\subsection{Individual Secrecy in Shannon's Ciphering System}
\begin{figure} [h]
\begin{center}
\includegraphics[scale=0.7]{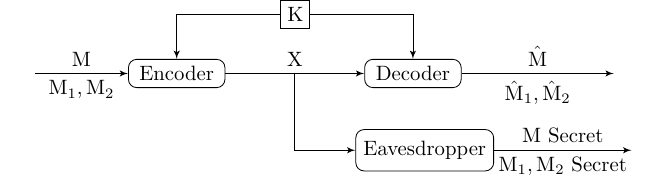}
\caption{Shannon's Cipher System}
\label{Fig:SCS}
\end{center}
\end{figure}

In this subsection, we will use Shannon's ciphering system to show why addressing individual secrecy with respect to different messages might be misleading, and that it is more consistent to interpret individuality with respect to different transmission flows. We consider the scenario given by Figure~\ref{Fig:SCS}. Shannon studied this model under the following secrecy constraint: 
\begin{equation}
\mathrm{\mathbb{I}(M;X)}= 0. \label{Equ:ShaSec}
\end{equation} 
He proved that this requirement is achieved if $\mathrm{\mathbb{H}(M) {\color{red}\leq} \mathbb{H} (K)}$, where $\mathrm{K}$ is the secret key shared between the transmitter and the receiver. In practical, it is hard to fulfill this condition because secret keys are usually shorter than the message. Now assume that we have a secret key such that $\mathrm{\mathbb{H}(K) = \frac{1}{2} \mathbb{H}(M)}$. We can construct the following coding strategy. First, we divide $\mathrm{M}$ into two messages $\mathrm{M}_1$ and $\mathrm{M}_2$, such that $\mathrm{\mathbb{H}(M_1) = \mathbb{H}(M_2) = \mathbb{H}(K)}$. We then construct a new secret key $\tilde{\mathrm{K}}$ by concatenating $\mathrm{K}$ and $\mathrm{M}_1$. Now the encoder outputs $\mathrm{X = M \otimes \tilde{K}}$, which is equivalent to the concatenation of $\mathrm{M_1 \otimes K}$ and $\mathrm{M_2 \otimes M_1}$. The decoder works in the following order, it first extracts $\mathrm{\hat{M}_1}$ from the first part of $\mathrm{X}$ by \textit{Xoring} it with the shared secret key $\mathrm{K}$, then it use $\mathrm{\hat{M}_1}$ to extract $\mathrm{\hat{M}}_2$ from the second part of $\mathrm{X}$. \newline
\indent Using this technique, we can overcome the problem of short secret key, however we need to understand the drawbacks of such technique. Aside form the problem of error progression that arises form using the estimated $\mathrm{\hat{M}_1}$ to decode $\mathrm{M}_2$, this technique does not fulfill the secrecy constraint in \eqref{Equ:ShaSec}. However, it fulfill the following individual secrecy constraint:
\begin{equation}
\mathrm{\mathbb{I}(M_1;X) + \mathbb{I}(M_2;X)} = 0. \label{Equ:ShaIndSec}
\end{equation} 
In general, we can extend this coding technique for short keys with smaller entropy by dividing the message $\mathrm{M}$ into smaller messages of the same entropy as the given key as $\mathrm{M} = \prod_{i=1}^L \mathrm{M}_i$. We can show that, the previous technique grants a certain secrecy level such that, the sum of the leakage of the small messages to the eavesdropper is small.\newline
\indent The difference between the two secrecy measures in the previous example is related to how to address the secrecy of information transmitted to a single user; whether it should be protected as a one big entity or it can be divided into smaller parts, where each part is protected separately. This issue is identical to the problem of identifying the individual secrecy and whether individuality means different users or different messages. That is why, we preferred the individual secrecy constraint in \eqref{Equ:IndCond} because it requires the whole information transmitted to a certain user to be protected as one big entity. In our opinion, this is a more consistent and meaningful notation.   

\subsection{Secrecy Capacity Regions: Joint Vs Individual}
In this subsection, we will try to highlight the differences between the joint and the individual secrecy criteria. To do so, we will compare the secrecy capacity region of both criteria for some special cases. Before we discuss these results, we need to introduce the following lemma.
\begin{Lemma}
Let $Q(y,z \arrowvert x)$ be a discrete memoryless BC and assume that $\mathrm{Y}$ is less noisy than $\mathrm{Z}$. Consider two independent random variables $\mathrm{M}$ and $\mathrm{W}$, such that $\mathrm{(M,W)} - \mathrm{X}^n - (\mathrm{Y}^n, \mathrm{Z}^n)$ forms a Markov chain. Then the following holds: $\mathrm{ \mathbb{I}(M;Y}^n \arrowvert \mathrm{W})\geq \mathrm{ \mathbb{I}(M;Z}^n \arrowvert \mathrm{W})$. 
\label{Prop:LessN}
\end{Lemma}
\begin{IEEEproof}
The proof uses a combination of standard techniques from \cite{Csis,LiangITS} and is given in Appendix \ref{App:LessN} for completeness.
\end{IEEEproof}
$ $ \newline
\indent In the first scenario, we consider a class of less noisy wiretap BC as in Figure \ref{Fig:BCWS}, where the eavesdropper is less noisy than the two legitimate receivers. We also modify the model such that, we only have the two individual confidential messages $\mathrm{M}_1$ and $\mathrm{M}_2$, without the common message $\mathrm{M}_c$ and the common confidential message $\mathrm{M}_0$. Thus, the joint secrecy conditions in \eqref{Equ:JoiCond} change to $\mathbb{I}(\mathrm{M}_1;\mathrm{Z}^n \arrowvert \mathrm{M}_2) \leq \tau_{1n}$ and $\mathbb{I}(\mathrm{M}_2;\mathrm{Z}^n \arrowvert \mathrm{M}_1) \leq \tau_{2n}$, while the individual secrecy conditions in \eqref{Equ:IndCond} change to $\mathbb{I}(\mathrm{M}_1;\mathrm{Z}^n) \leq \tau_{1n}$ and $\mathbb{I}(\mathrm{M}_2;\mathrm{Z}^n) \leq \tau_{2n}$.
\begin{mytheo}
\label{Ther:2}
Consider a wiretap BC with message cognition, where the eavesdropper $\mathrm{Z}$ is less noisy than the two legitimate receivers $\mathrm{Y_1}$ and $\mathrm{Y_2}$, i.e. $\mathrm{Z \succeq Y_1}$ and $\mathrm{Z \succeq Y_2}$. Then the joint secrecy capacity region is empty, while the individual secrecy capacity region is given by the set of all rate pairs $(R_1,R_2) \in \mathbb{R}_+^2$ that satisfy
\begin{equation}
R_1 = R_2 \leq \min \Big[\mathrm{\mathbb{I}(X;Y_1), \mathbb{I}(X;Y_2)} \Big]. 
\label{Equ:ICap2} 
\end{equation} 
\end{mytheo}
\begin{IEEEproof}
We start with the individual secrecy capacity region. The proof of the achievability is based on interpreting each individual message as a secret key for the other one. The encoder constructs the \textit{Xored} message $\mathrm{M}_\otimes$ by \textit{Xoring} the corresponding elements of $\mathrm{M}_1$ and $\mathrm{M}_2$ as follows:
\begin{equation*}
m_\otimes = m_1 \otimes m_2.
\end{equation*}
In order to transmit a message pair $(m_1,m_2)$, the encoder generates the sequence $\mathrm{X}^n (m_\otimes)$, then transmits it to both receivers. The problem simplifies to a multicast problem and reliable transmission is only guaranteed by the condition in \eqref{Equ:ICap2}. Each legitimate receiver decodes the \textit{Xored} message $\mathrm{M}_\otimes$ then uses the side information to extract it is own message. On the other hand, the eavesdropper can not extract any information about $\mathrm{M}_1$ and $\mathrm{M}_2$, although it can correctly decode $\mathrm{M}_\otimes$, because $\mathbb{I}\mathrm{(M_\otimes,M_1)} = 0$ and $\mathbb{I}\mathrm{(M_\otimes,M_2)} = 0$.\newline
\indent Now for the converse, using Lemma~\ref{Prop:LessN}, we will show that, if $\mathrm{Z}$ is less noisy than both $\mathrm{Y}_1$ and $\mathrm{Y}_2$, the two rates $R_1$ and $R_2$ are equal. Let $\epsilon_n$ and $\tau_n= \max(\tau_{1n},\tau_{2n})$ be two sequences, such that as $n \rightarrow \infty$, $\epsilon_n$ and $\tau_n \rightarrow 0$, we have 
\begin{align}
R_1 &\overset{(a)}{\leq} \frac{1}{n} \mathrm{\mathbb{I}(M_1;Y}_1^n \arrowvert \mathrm{M}_2) + \gamma_1(\epsilon_n) \nonumber\\
&\leq \frac{1}{n} \mathrm{\mathbb{I}(M_1 M_2;Y}_1^n) + \gamma_1(\epsilon_n) \nonumber\\
&\overset{(b)}{\leq} \frac{1}{n} \Big[ \mathrm{\mathbb{I}(M_1 M_2;Y}_1^n) - \mathrm{\mathbb{I} (M_1;Z}^n) \Big] + \gamma_1(\epsilon_n,\tau_{n})  \nonumber\\
&= \frac{1}{n} \Big[ \mathrm{\mathbb{I}(M_1 M_2;Y}_1^n) - \mathrm{\mathbb{I}(M_1 M_2;Z}^n) + \mathrm{\mathbb{I} (M_2;Z}^n \arrowvert \mathrm{M_1}) \Big] \nonumber\\
&\qquad + \gamma_1(\epsilon_n,\tau_{n})  \nonumber\\
&\overset{(c)}{\leq} \frac{1}{n}\mathrm{\mathbb{I} (M_2;Z}^n \arrowvert \mathrm{M_1}) + \gamma_1(\epsilon_n, \tau_n) \nonumber\\ 
&\overset{(d)}{\leq} R_2+\gamma_1(\epsilon_n,\tau_{n}), 
\label{Equ:IConv2_1}
\end{align}
where $(a)$ follows from Fano's inequality as $\gamma_1(\epsilon_n) = 1/n + \epsilon_n R_1$; $(b)$ follows from \eqref{Equ:IndCond}, when $\mathrm{M}_0 = \emptyset$ and $\gamma_1(\epsilon_n, \tau_{n}) = (1+\tau_{n})/n + \epsilon_n R_1$; $(c)$ follows from Lemma~\ref{Prop:LessN} because $\mathrm{Z} \succeq \mathrm{Y}_1$, which implies that $ \mathrm{\mathbb{I}(M_1 M_2;Y}_1^n) - \mathrm{\mathbb{I}(M_1 M_2;Z}^n) \leq 0$ and $(d)$ follows because $R_2 \geq 1/n \mathrm{\mathbb{I} (M_2;Z}^n \arrowvert \mathrm{M_1})$. If we let $\gamma_2(\epsilon_n,\tau_{2n}) = (1+\tau_{2n})/n + \epsilon_n R_2$ and follow the same steps we can derive a similar bound for $R_2$ as follows: 
\begin{equation}
R_2 \leq R_1 + \gamma_2(\epsilon_n,\tau_{2n}). \label{Equ:IConv2_2}
\end{equation}
Now in order to finalize our converse we need to highlight the standard upper bound for reliable transmission for each receiver given by:
\begin{equation}
R_1 \leq \mathbb{I}(\mathrm{X;Y_1}) \hspace{0.5cm} \text{and} \hspace{0.5cm}
R_2 \leq \mathbb{I}(\mathrm{X;Y_2}).
\label{Equ:IConv2_3}
\end{equation}
Finally, if we take the limit as $n \rightarrow \infty$ for \eqref{Equ:IConv2_1}, \eqref{Equ:IConv2_2}, such that $\gamma_1(\epsilon_n,\tau_n)$ and $\gamma_2( \epsilon_n,\tau_n) \rightarrow 0$, Our converse for the individual secrecy capacity region in \eqref{Equ:ICap2} is complete. \newline
\indent Now, we turn to the other half of the theorem that indicates that the joint secrecy capacity region is empty if the eavesdropper is less noisy than the two legitimate receivers. The proof is based on Lemma~\ref{Prop:LessN} and is a direct consequence of \cite[Proposition 3.4]{Bloch} and \cite{Csis} as follows:
\begin{align}
R_1 &\overset{(a)}{\leq} \frac{1}{n} \mathrm{\mathbb{I}(M_1;Y}_1^n \arrowvert \mathrm{M}_2) + \gamma_1(\epsilon_n) \nonumber\\
&\overset{(b)}{\leq} \frac{1}{n} \Big[\mathrm{\mathbb{I}(M_1;Y}_1^n \arrowvert \mathrm{M}_2) - \mathrm{\mathbb{I}(M_1;Z}^n \arrowvert \mathrm{M}_2) \Big] + \gamma_1(\epsilon_n,\tau_{n})  \nonumber\\
&\overset{(c)}{\leq} \gamma_1(\epsilon_n, \tau_n).
\label{Equ:IConv2_4}
\end{align}
where $(a)$ follows from Fano's inequality; $(b)$ follows from \eqref{Equ:JoiCond}, for $\mathrm {M}_0 = \emptyset$; while $(c)$ follows from Lemma~\ref{Prop:LessN} because $\mathrm{Z} \succeq \mathrm{Y}_1$, which implies that $\mathrm{\mathbb{I}(M_1;Y}_1^n \arrowvert \mathrm{M}_2) \leq \mathrm{ \mathbb{I}(M_1;Z}^n \arrowvert \mathrm{M}_2)$. Similarly, we have for $R_2$ the following 
\begin{equation} 
R_2 \leq \gamma_2(\epsilon_n,\tau_{n}). \label{Equ:IConv2_5}
\end{equation}
Now if we take the limit as $n \rightarrow \infty$ for \eqref{Equ:IConv2_4}, \eqref{Equ:IConv2_5},  such that $\gamma_1(\epsilon_n,\tau_n)$ and $\gamma_2( \epsilon_n,\tau_n) \rightarrow 0$, we have $R_1 = R_2 = 0$. This implies that the joint secrecy capacity region for this scenario is empty. 
\end{IEEEproof} 
\begin{Remark}
The previous result was established for wiretap BC with receiver side information, where the legitimate receivers $\mathrm{Y}_1$ and $\mathrm{Y}_2$ are degraded from the eavesdropper $\mathrm{Z}$ in \cite{Aydin2}.
\end{Remark}
\indent In the next scenario, we will continue with the previous model, where we discuss the wiretap BC in Figure \ref{Fig:BCWS} with only $\mathrm{M}_1$ and $\mathrm{M}_2$. However, we will investigate a different class of less noisy channels, where the two legitimate receivers $\mathrm{Y_1}$ and $\mathrm {Y_2}$ are less noisy than the eavesdropper $\mathrm{Z}$.
\begin{mytheo}
Consider a wiretap BC with message cognition, where the two legitimate receivers $\mathrm{Y_1}$ and $\mathrm {Y_2}$ are less noisy than the eavesdropper $\mathrm{Z}$, i.e. $\mathrm{Y_1 \succeq Z}$ and ${\color{red}\mathrm{Y_2 \succeq Z}}$. Then the joint secrecy capacity region is given by the set of all rate pairs $(R_1,R_2) \in \mathbb{R}_+^2$, such that
\begin{equation}
\begin{split}
R_1 &\leq \mathrm{\mathbb{I} (X;Y_1) - \mathbb{I}(X;Z)} \\
R_2 &\leq \mathrm{\mathbb{I} (X;Y_2) - \mathbb{I}(X;Z)}.  
\end{split} 
\label{Equ:JCapEx2} 
\end{equation}
While, the individual secrecy capacity region for the same scenario is given by the set of all rate pairs $(R_1,R_2) \in \mathbb{R}_+^2$ that satisfy
\begin{equation}
\begin{split}
R_1 &\leq \min \Big[\mathrm{\mathbb{I} (X;Y_1) - \mathbb{I}(X;Z)} +R_2 \mathrm{\text{ , }\mathbb{I} (X;Y_1)} \Big]  \\
R_2 &\leq \min \Big[\mathrm{\mathbb{I} (X;Y_2) - \mathbb{I}(X;Z)} +R_1 \mathrm{\text{ , }\mathbb{I} (X;Y_2)} \Big]. 
\end{split} 
\label{Equ:ICapEx2} 
\end{equation}
\end{mytheo} 
\begin{Remark}
Since the class of less noisy channels includes the class of physically and stochastically degraded channels, the previous theorem generalizes the secrecy capacity regions established in \cite{Aydin2}, for the wiretap BC with message cognition, where the eavesdropper is degraded from both legitimate receivers. 
\end{Remark}
\begin{IEEEproof}
We will only give a sketch for the ideas of the proof as we will present a detailed proof in the next sections for a more general case. The achievability of the joint secrecy region follows from technique of random coding with product structure as in \cite{Csis}, while the achievability of the individual secrecy region combines the techniques of wiretap random coding along with Shannon's one time pad cipher system used in Theorem~\ref{Ther:2}, where the ciphered message is used as a part of the randomization index needed for the wiretap random coding. This encoding scheme was first introduced in \cite{Ind1}. On the other hand, the converse for the joint secrecy region follows using the standard techniques and procedures used in \cite{Gamal} for less noisy channels. While the converse for the individual secrecy region follows by adapting those techniques to the individual secrecy constraint.
\end{IEEEproof} \vspace{0.1cm}
\indent Differently from the previous two scenarios, where the joint and the individual secrecy criteria lead to different capacity regions, in the next example, we will investigate a scenario where the two secrecy criteria are equivalent. Consider a wiretap BC as in Figure \ref{Fig:BCWS}, where we only have the common message $\mathrm{M}_c$ and the common confidential message $\mathrm{M}_0$. One can easily conclude by comparing the requirements of the joint secrecy and the individual secrecy in \eqref{Equ:JoiCond} and \eqref{Equ:IndCond} when $\mathrm{M}_1 = \mathrm{M}_2 = \emptyset$, that the two secrecy criteria are the same. Again, we will focus on a class of less noisy channels, where one of the legitimate receivers is less noisy than the other one, while the relation to the eavesdropper is arbitrary.
\begin{mytheo}
The joint and individual secrecy capacity region for the wiretap BC with a common message and one confidential message, if one of the legitimate receivers is less noisy than the other one $(\mathrm{Y_1} \succeq \mathrm {Y_2})$, is the set of all rates $(R_c,R_0) \in \mathbb{R}_+^2$ that satisfy
\begin{align}
R_c &\leq \min \Big[\mathrm{\mathbb{I}(U;Y_2),\mathbb{I}(U;Z)} \Big] \nonumber \\
R_0 &\leq \mathrm{\mathbb{I}(V;Y_2 \arrowvert U) - \mathbb{I}(V;Z \arrowvert U)} \label{Equ:Cap3} 
\end{align}
for some $\mathrm{(U,V,X)}$, such that $\mathrm{U- V - X - (Y_1, Y_2, Z)}$ forms a Markov chain. Further it suffices to have $\arrowvert \mathcal{U} \arrowvert \leq \arrowvert \mathcal{X} \arrowvert + 3$ and $\arrowvert \mathcal{V} \arrowvert \leq \arrowvert \mathcal{X} \arrowvert^2 + 4\arrowvert \mathcal{X} \arrowvert + 3$.
\end{mytheo}
\begin{IEEEproof}
The achievability follows from the straightforward extension of the Csisz\'{a}r-K\"{o}rner results in \cite{Csis}, leading to the following lower bounds:
\begin{align}
R_c &\leq \min \Big[\mathrm{\mathbb{I}(U;Y_1),\mathbb{I}(U;Y_2),\mathbb{I}(U;Z)}\Big] \nonumber \\
R_0 &\leq \mathrm{\mathbb{I}(V;Y_1 \arrowvert U) - \mathbb{I}(V;Z \arrowvert U)} \nonumber \\
R_0 &\leq \mathrm{\mathbb{I}(V;Y_2 \arrowvert U) - \mathbb{I}(V;Z \arrowvert U)}. \label{Equ:Ach3_1} 
\end{align}
Since $\mathrm{Y}_1 \succeq \mathrm{Y}_2$, which implies that $\mathrm{ \mathbb{I}(U; Y_2) \leq \mathbb{I}(U;Y_1)}$ and $\mathrm{\mathbb{I}(V;Y_2 \arrowvert U) \leq \mathbb{I}(V;Y_1 \arrowvert U})$. Substituting these two relations in \eqref{Equ:Ach3_1} leads the achievability of the region in \eqref{Equ:Cap3}. On the other hand, the converse follows directly using the standard techniques in \cite[Theorem 1]{Csis}.
\end{IEEEproof} 
\subsection{Discussion}
The previous examples are very helpful in understanding the differences between the joint and individual secrecy criteria. They also helps in capturing the advantages and disadvantages of each one. This can be summarized in the following points: \vspace{0.05cm} \newline
\indent 1. Any code that satisfies the joint secrecy criterion will also satisfy the individual one as well. This advocates the fact that the individual secrecy is a less conservative secrecy measure as compared to the joint one. \vspace{0.1cm} \newline
\indent 2. The individual secrecy criterion provides a larger capacity region as compared to the joint one. Even if the joint capacity region is zero, the individual criterion can provide an non vanishing achievable rate. This increase in the rate comes from the usage of secret key encoding in addition to the standard random wiretap encoding. That is why the value of this increase is directly proportional with the size of the individual messages cf. \eqref{Equ:JCapEx2} and \eqref{Equ:ICapEx2}. \vspace{0.1cm} \newline
\indent 3. The joint secrecy criterion is a very conservative secrecy measure. Even if one of the confidential messages is revealed to the eavesdropper in a genie-aided way, the other message is still protected as follows:
\begin{align}
\mathbb{I}(\mathrm{M}_1;\mathrm{Z}^n \mathrm{M}_2) &= \mathbb{I}(\mathrm{M_1;M_2}) + \mathbb{I} (\mathrm{M}_1;\mathrm{Z}^n \arrowvert \mathrm{M}_2) \nonumber\\
&\overset{(a)}{=} \mathbb{I} (\mathrm{M}_1;\mathrm{Z}^n \arrowvert \mathrm{M}_2) \leq \tau_n,
\end{align}
where $(a)$ follows because $\mathrm{M}_1$ and $\mathrm{M}_2$ are independent. The previous equation shows that the leakage of $\mathrm{M}_1$ to the eavesdropper when $\mathrm{M}_2$ is revealed to it is still small.\vspace{0.1cm} \newline
\indent 4. On the other hand, the individual secrecy criterion is based on the mutual trust between the legitimate receivers. Thus if one of the messages is compromised, this might also affects the secrecy of the other one. In order to understand this property, imagine that in the previous two examples, $\mathrm{M}_2$ was revealed to the eavesdropper as follows:
\begin{align}
\mathbb{I}(\mathrm{M}_1;\mathrm{Z}^n \mathrm{M}_2) &= \mathbb{H}(\mathrm{M}_1) - \mathbb{H}(\mathrm {M_1} \arrowvert \mathrm{Z}^n \mathrm{M}_2).
\end{align} 
In the first scenario, where the eavesdropper $\mathrm{Z}$ is less noisy than the two legitimate receivers, the term $\mathbb{H}(\mathrm {M_1} \arrowvert \mathrm{Z}^n \mathrm{M}_2)$ will vanish. This is because the eavesdropper can correctly decode $\mathrm{M}_\otimes$, then using the secret key $\mathrm{M}_2$, it can extract $\mathrm{M}_1$ as well. This implies that $\mathrm{M}_1$ is fully leaked to the eavesdropper when $\mathrm{M}_2$ is revealed to it. However, in the second scenario, the situation is a little bit different. This is because the term $\mathbb{H}(\mathrm {M_1} \arrowvert \mathrm{Z}^n \mathrm{M}_2)$ does not vanish, yet it is smaller than $\mathbb{H} (\mathrm{M}_1)$. This means that a part of $\mathrm{M}_1$ is leaked to the eavesdropper up on revealing $\mathrm{M}_2$. The size of this part depends on how much the eavesdropper can infer using its received signal $\mathrm{Z}^n$ and $\mathrm{M}_2$. \vspace{0.1cm}\newline
\indent 5. The preference in choosing among the two secrecy criteria is a trade of between conservative secrecy measure and a larger capacity region and the decision should always be based on whether the legitimate receivers can trust one another or not.
\section{The Joint Secrecy Capacity Region}
In this section, we investigate the joint secrecy criterion for the general model of the wiretap BC with degraded message sets and message cognition given by Figure~\ref{Fig:BCWS}.
\label{Sec:BCJS}
\subsection{Achievable Rate Region}
\begin{Prop}
An achievable joint secrecy rate region for the wiretap BC with degraded message sets and message cognition is given by the set of all rate quadruples $(R_c,R_0,R_1,R_2) \in \mathbb{R}_+^4$ that satisfy
\begin{align}
R_c &\leq \mathrm{\mathbb{I}(U;Z)} \nonumber \\
R_0 + R_1 &\leq \mathrm{\mathbb{I}(V_0 V_1;Y_1 \arrowvert U) - \mathbb{I}(V_0 V_1; Z \arrowvert U)} \nonumber \\
R_0 + R_2 &\leq \mathrm{\mathbb{I}(V_0 V_2;Y_2 \arrowvert U) - \mathbb{I}(V_0 V_2; Z \arrowvert U)} \nonumber \\
R_c + R_0 + R_1 &\leq \mathrm{\mathbb{I}(V_0 V_1;Y_1) - \mathbb{I}(V_0 V_1; Z \arrowvert U)} \nonumber \\
R_c + R_0 + R_2 &\leq \mathrm{\mathbb{I}(V_0 V_2;Y_2) - \mathbb{I}(V_0 V_2; Z \arrowvert U)} \nonumber \\
2R_0 + R_1 + R_2 &\leq \mathrm{\mathbb{I}(V_0 V_1;Y_1 \arrowvert U) + \mathbb{I} (V_0 V_2;Y_2 \arrowvert U)} \nonumber\\
\mathrm{-\mathbb{I}(V_1;V_2 \arrowvert} &\mathrm{V}_0) \mathrm{-\mathbb{I} (V_0V_1V_2;Z \arrowvert U) - \mathbb{I}(V_0;Z \arrowvert U)} \nonumber\\
R_c + 2R_0 + R_1 + R_2 &\leq \mathrm{\mathbb{I}(V_0 V_1;Y_1) + \mathbb{I} (V_0 V_2;Y_2 \arrowvert U) }\nonumber\\
\mathrm{-\mathbb{I}(V_1;V_2 \arrowvert} &\mathrm{V}_0) \mathrm{-\mathbb{I}(V_0V_1V_2;Z \arrowvert U) - \mathbb{I}(V_0;Z \arrowvert U)} \nonumber\\
R_c + 2R_0 + R_1 + R_2 &\leq \mathrm{\mathbb{I}(V_0 V_1;Y_1 \arrowvert U) + \mathbb{I} (V_0 V_2;Y_2) } \nonumber\\
\mathrm{-\mathbb{I}(V_1;V_2 \arrowvert} &\mathrm{V}_0) \mathrm{-\mathbb{I}(V_0V_1V_2;Z \arrowvert U) - \mathbb{I}(V_0;Z \arrowvert U)}
\label{Equ:Rates} 
\end{align}
for random variables with joint probability distribution $Q(u)$ $Q(v_0 \arrowvert u)$ $Q(v_1,v_2 \arrowvert v_0)$ $Q(x\arrowvert v_1,v_2)$ $Q(y_1, y_2, z \arrowvert x)$, such that $\mathrm{U - V_0 - (V_1,V_2) - X - (Y_1,Y_2,Z)}$ forms a Markov chain.
\label{Lem:Lemma1}
\end{Prop}
\begin{IEEEproof}
The proof combines the principle of superposition random coding \cite{Csis} in addition to the usage of Marton coding for secrecy as in \cite{Gamal}, where strong secrecy is achieved as in \cite{Bloch_Resol,Kram2,MoritzSS}. \vspace{0.2cm} \newline
\indent \underline{\textbf{1. Message sets:}} We consider the following sets: The set of common messages $\mathcal{M}_c = \llbracket 1, 2^{nR_c} \rrbracket$, the set of confidential common messages $\mathcal{M}_0 = \llbracket 1, 2^{nR_0} \rrbracket$, two sets of confidential individual messages $\mathcal{M}_1 = \llbracket 1, 2^{nR_1} \rrbracket$ and $\mathcal{M}_2 = \llbracket 1, 2^{nR_2} \rrbracket$, three sets of randomization messages for secrecy $\mathcal{M}_r =\llbracket 1, 2^{nR_r} \rrbracket$, $\mathcal{M}_{r_1} = \llbracket 1, 2^{nR_{r_1}} \rrbracket$ and $\mathcal{M}_{r_2} = \llbracket 1, 2^{nR_{r_2}} \rrbracket$, finally two additional sets $\mathcal{M}_{t_1} = \llbracket 1, 2^{nR_{t_1}} \rrbracket$ and $\mathcal{M}_{t_2} = \llbracket 1, 2^{nR_{t_2}} \rrbracket$ needed for the construction of Marton coding. Additionally we use $\mathcal{M} = \mathcal{M}_0 \times \mathcal{M}_1 \times \mathcal{M}_2$ to abbreviate the set of all confidential messages. \vspace{0.2cm} 

\underline{\textbf{2. Random Codebook $\mathcal{C}^s_n$:}} Fix an input distribution $Q(u,v_0,v_1,v_2,x)$. Construct the codewords $u^n(m_c)$ for $m_c \in \mathcal{M}_c$ by generating symbols $u_i (m_c)$ with $i \in \llbracket 1, n \rrbracket$ independently according to $Q(u)$. For every $u^n(m_c)$, generate codewords $v_0^n (m_c,m,m_r)$ for $m \in \mathcal{M}$ and $m_r \in \mathcal{M}_r$ by generating symbols $v_{0_i}(m_c,m,m_r)$ independently at random according to $Q(v_0 \arrowvert u_i(m_c))$. Next, for each $v_0^n (m_c,m,m_r)$ generate the codewords $v_1^n(m_c,m,m_r,m_{r_1},m_{t_1})$ and $v_2^n(m_c,m,m_r,m_{r_2},m_{t_2})$ for $m_{r_1} \in \mathcal{M}_{r_1}$, $m_{r_2} \in \mathcal{M}_{r_2}$, $m_{t_1} \in \mathcal{M}_{t_1}$ and $m_{t_2} \in \mathcal{M}_{t_2}$ by generating symbols $v_{1_i}(m_c,m,m_r,m_{r_1},m_{t_1})$ and $v_{2_i}(m_c,m,m_r,m_{r_2},m_{t_2})$ independently at random according to $Q (v_1 \arrowvert v_{0_i}(m_c,m,m_r))$ and $Q (v_2 \arrowvert v_{0_i}(m_c,m,m_r))$ respectively. \vspace{0.2cm} 

\underline{\textbf{3. Encoder $E$:}} Given a message pair $(m_c,m)$, where $m = (m_0, m_1, m_2)$, the transmitter chooses three randomization messages $m_r$, $m_{r_1}$ and $m_{r_2}$ uniformly at random from the sets $\mathcal{M}_r$, $\mathcal{M}_{r_1}$ and $\mathcal{M}_{r_2}$ respectively. Then, it finds a pair $(m_{t_1},m_{t_2})$ such that $v_1^n (m_c,m,m_r,m_{r_1}, m_{t_1})$ and $v_2^n (m_c,m,m_r,m_{r_2},m_{t_2})$ are jointly typical. Finally, it generates a codeword $x^n$ independently at random according to $\prod_{i=1}^n Q(x_i \arrowvert v_{1_i},v_{2_i})$ and transmits it. \vspace{0.2cm} 
 
\underline{\textbf{4. First Legitimate Decoder $\varphi_1$:}} Given $y_1^n$ and its own message $m_2$, outputs $(\hat{m}_c,\hat{m}_0,\hat{m}_1,\hat{m}_r)$; if they are the unique messages, such that $u^n(\hat{m}_c)$, $v_0^n(\hat{m}_c,\hat{m},\hat{m_r})$, $v_1^n(\hat{m}_c, \hat{m}, \hat{m}_r, \hat{m}_{r_1},\hat{m}_{t_1})$ and $y_1^n$ are jointly typical, for some $(\hat{m}_{r_1}, \hat{m}_{t_1}) \in \mathcal{M}_{r_1} \times \mathcal{M}_{t_1}$, where $\hat{m}= (\hat{m}_0, \hat{m}_1,m_2)$. Otherwise declares an error. \vspace{0.2cm}

\underline{\textbf{5. Second Legitimate Decoder $\varphi_2$:}} Given $y_2^n$ and its own message $m_1$, outputs $(\tilde{m}_c,\tilde{m}_0,\tilde{m}_2,\tilde{m}_r)$; if they are the unique messages, such that $u^n(\tilde{m}_c) $, $v_0^n (\tilde{m}_c,\tilde{m},\tilde{m}_r)$, $v_2^n (\tilde{m}_c, \tilde{m},\tilde{m}_r,\tilde{m}_{r_2},\tilde{m}_{t_2})$ and $y_2^n$ are jointly typical, for some $(\tilde{m}_{r_2},\tilde{m}_{t_2}) \in \mathcal{M}_{r_2} \times \mathcal{M}_{t_2}$, where $\tilde{m}= (\tilde{m}_0,m_1,\tilde{m}_2)$. Otherwise declares an error. \vspace{0.2cm}

\underline{\textbf{6. Third Eavesdropper Decoder $\varphi_3$:}} Given $z^n$, outputs $\check{m}_c$; if it is the unique message, such that $u^n(\check{m}_c)$ and $z^n$ are jointly typical. Otherwise declares an error. \vspace{0.2cm} 

\underline{\textbf{7. Reliability Analysis:}} We define the average error probability of this scheme as
\begin{align*}
\hat{P}_e &(\mathcal{C}_n) \triangleq \mathbb{P} \big[(\hat{\mathrm{M}}_c, \hat{\mathrm{M}}_0,\hat{\mathrm{M}}_1,\hat{\mathrm{M}}_r) \neq (\mathrm{M}_c, \mathrm{M}_0, \mathrm{M}_1,\mathrm{M}_r) \text{ or }   \nonumber\\
&(\tilde{\mathrm{M}}_c, \tilde{\mathrm{M}}_0, \tilde{\mathrm{M}}_2, \tilde{\mathrm{M}}_r) \neq (\mathrm{M}_c, \mathrm{M}_0,\mathrm{M}_2, \mathrm{M}_r) \text{ or } \check{\mathrm{M}}_c \neq \mathrm{M}_c \big].
\end{align*}
We then observe that $\hat{P}_e (\mathcal{C}_n) \geq P_e (\mathcal{C}_n)$, cf. \eqref{Equ:ErrorProb}. Using the standard analysis of random coding we can prove that for a sufficiently large $n$, with high probability $\hat{P}_e (\mathcal{C}_n) \leq \epsilon_n$ if 
\begin{align}
R_c &\leq \mathrm{\mathbb{I}(U;Z)} - \delta_n(\epsilon_n) \nonumber\\ 
\sq{R_{t_1} + R_{t_2}} &\geq \mathrm{\mathbb{I} (V_1;V_2 \arrowvert V_0)} + \delta_n (\epsilon_n) \nonumber \\ 
\sq{R_0 + R_1 + R_r + R_{r_1} + R_{t_1}} &\leq \mathrm{ \mathbb{I} (V_0 V_1; Y_1 \arrowvert U)} - \delta_n (\epsilon_n)\nonumber \\
\sq{R_0 + R_2 + R_r + R_{r_2} + R_{t_2}} &\leq \mathrm{ \mathbb{I} (V_0 V_2; Y_2 \arrowvert U)} - \delta_n (\epsilon_n) \nonumber \\
\sq{R_c + R_0 + R_1 + R_r + R_{r_1} + R_{t_1}} &\leq \mathrm{ \mathbb{I} (V_0 V_1; Y_1)} - \delta_n (\epsilon_n) \nonumber \\  
\sq{R_c + R_0 + R_2 + R_r + R_{r_2} + R_{t_2}} &\leq \mathrm{ \mathbb{I} (V_0 V_2; Y_2)} - \delta_n (\epsilon_n). \label{Equ:RelRate}
\end{align}
The validity of \eqref{Equ:RelRate} follows from the product structure of the codebook, the full cognition of the individual messages at the legitimate receivers and the principle of indirect decoding used in \cite{Gamal}. In addition, the second constraint follows due to Marton coding technique, where the summation of $R_{t_1}$ and $R_{t_2}$ should be greater than $\mathrm{\mathbb{I} (V_1;V_2 \arrowvert V_0)}$ to guarantee the existence of a typical pair $(v_1^n,v_2^n)$. \vspace{0.2cm}

\underline{\textbf{8. Secrecy Analysis:}} Our secrecy analysis is based on different strong secrecy techniques as in \cite{Bloch_Resol,Kram2,MoritzSS}. We start by identifying all the virtual channels that exist between the confidential messages and the eavesdropper. Based on the codebook structure, we can define four possible channels as follow: $\mathrm{Q}_1: \mathcal{V}_0 \rightarrow \mathcal{P}(\mathcal{Z})$, $\mathrm{Q}_2: \mathcal{V}_0 \times \mathcal{V}_1  \rightarrow \mathcal{P}(\mathcal{Z})$, $\mathrm{Q}_3: \mathcal{V}_0 \times \mathcal{V}_2  \rightarrow \mathcal{P}(\mathcal{Z})$ and $\mathrm{Q}_4: \mathcal{V}_0 \times \mathcal{V}_1 \times \mathcal{V}_2 \rightarrow \mathcal{P}(\mathcal{Z})$. According to \cite{Kram2}, in order to fulfill the joint strong secrecy criterion in \eqref{Equ:MJoiSec}, we need to make sure that the randomization rate in the input sequence to each of these virtual channels is at least equivalent to the mutual information between the channel input and the eavesdropper. Thus, for a sufficiently large $n$ and $\tau_n >0$, the joint secrecy constraints given in \eqref{Equ:MJoiSec} is with high probability smaller than $\tau_n$, if
\begin{align}
R_r &\geq \mathrm{\mathbb{I}(V_0;Z \arrowvert U)} +\delta_n(\tau_n) \nonumber \\
R_r + R_{r_1} + R_{t_1} &\geq  \mathrm{\mathbb{I}(V_0 V_1;Z \arrowvert U)} +\delta_n(\tau_n) \nonumber \\
R_r + R_{r_2} + R_{t_2} &\geq  \mathrm{\mathbb{I}(V_0 V_2;Z \arrowvert U)} +\delta_n(\tau_n) \nonumber \\
R_r + R_{r_1} + R_{r_2} &\geq  \mathrm{\mathbb{I}(V_0 V_1 V_2;Z \arrowvert U)} +\delta_n(\tau_n).
\label{Equ:SecRate}
\end{align}    
\indent If we combine Eq. \eqref{Equ:RelRate} and Eq. \eqref{Equ:SecRate} then apply the Fourier-Motzkin elimination procedure, followed by taking the limit as $n \rightarrow \infty$, which implies that $\delta_n(\epsilon_n)$ and $\delta_n(\tau_n) \rightarrow 0$, we prove the achievability of any rate quadruple $(R_c,R_0,R_1,R_2)$ satisfying \eqref{Equ:Rates}.
\end{IEEEproof}

\subsection{Secrecy Capacity For A Class of More Capable Channels}
\begin{mytheo}
Consider a wiretap BC with degraded message sets and message cognition, where one of the legitimate receivers $\mathrm{Y_1}$ is more capable than the eavesdropper $\mathrm{Z}$, while the relation between the other legitimate receiver $\mathrm{Y_2}$ and the eavesdropper $\mathrm{Z}$ is arbitrary. Then, the joint secrecy capacity region is given by the set of all rate quadruples $(R_c,R_0,R_1,R_2) \in \mathbb{R}_+^4$ that satisfy
\begin{align}
R_c &\leq \mathrm{\mathbb{I} (U;Z)} \nonumber \\
R_0 + R_1 &\leq \mathrm{\mathbb{I}(X;Y_1\arrowvert U) - \mathbb{I}(X;Z \arrowvert U)} \nonumber \\
R_0 + R_2 &\leq \mathrm{\mathbb{I}(V;Y_2\arrowvert U) - \mathbb{I}(V;Z \arrowvert U)} \nonumber \\
R_c + R_0 + R_1 &\leq \mathrm{\mathbb{I}(X;Y_1) - \mathbb{I}(X;Z \arrowvert U)} \nonumber \\
R_c + R_0 + R_2 &\leq \mathrm{\mathbb{I}(V;Y_2) - \mathbb{I}(V;Z \arrowvert U)}
\label{Equ:Cap1} 
\end{align}
for some $\mathrm{(U,V,X)}$, such that $\mathrm{U - V - X - (Y_1, Y_2, Z)}$ forms a Markov chain. Further it suffices to have $\arrowvert \mathcal{U} \arrowvert \leq \arrowvert \mathcal{X} \arrowvert + 3$ and $\arrowvert \mathcal{V} \arrowvert \leq \arrowvert \mathcal{X} \arrowvert^2 + 4\arrowvert \mathcal{X} \arrowvert + 3$.
\label{Ther:1}
\end{mytheo}
\begin{IEEEproof}
The achievability is based on the principle of indirect decoding introduced in \cite{DMSNG} and its extension to secrecy scenarios discussed in \cite{Gamal}. It also follows directly from Proposition \ref{Lem:Lemma1}, by letting $\mathrm{V_2} = \emptyset$, $\mathrm{V_0 = V}$ and $\mathrm{V_1 = X}$  in \eqref{Equ:Rates}. This implies that the first legitimate receiver $\mathrm{Y}_1$ which is more capable than the eavesdropper $\mathrm{Z}$ finds its intended messages by direct decoding from $\mathrm{X}$, while the second legitimate receiver $\mathrm{Y}_2$ which has no stastical advantage over the eavesdropper $\mathrm{Z}$ finds its intended messages by indirect decoding from the auxiliary random variable $\mathrm{V}$. \newline
\indent For the converse, we start by modifying the joint secrecy constraint in \eqref{Equ:MJoiSec} to include the conditioning on the common message. For this we need the following lemma:
\begin{Lemma}
Consider two independent random variables $\mathrm{M}$ and $\mathrm{W}$, such that $\mathbb{H}( \mathrm{W} \arrowvert \mathrm{Z}^n) \leq \alpha$ and $\mathbb{I}(\mathrm{M};\mathrm{Z}^n) \leq \beta$, where $\alpha,\beta > 0$. Then, $\mathbb{I}(\mathrm{M}; \mathrm{Z}^n \arrowvert \mathrm{W}) \leq \alpha + \beta$ holds.
\label{Prop:CMC} 
\end{Lemma}
\begin{IEEEproof}
The proof is based on the properties of the entropy function and is given in Appendix \ref{App:CMC} for completeness.
\end{IEEEproof}
$ $ \newline
Since Eq. \eqref{Equ:RelComCod} implies that $\mathbb{H}(\mathrm{M}_c \arrowvert \mathrm{Z}^n) \leq {\color{red}n} \gamma_c(\epsilon_n)$ and Eq. \eqref{Equ:MJoiSec} implies that $\mathbb{I}(\mathrm{M_0 M_1 M_2}; \mathrm{Z}^n) \leq \tau_n$, we can use the previous lemma to reformulate the joint secrecy constraint for our scenario as:
\begin{equation}
\mathbb{I}(\mathrm{M_0 M_1 M_2}; \mathrm{Z}^n \arrowvert \mathrm{M}_c) \leq n\gamma_c(\epsilon_n) + \tau_n.  
\label{Equ:JoiSecCon}
\end{equation} 
\indent Now, we are ready to formulate our converse. First, we let $\mathrm{U}_i \triangleq (\mathrm{M}_c, \tilde{\mathrm{Z}}^{i+1})$, $\mathrm{K}_i^1 \triangleq \mathrm{Y}_1^{i-1}$, $\mathrm{K}_i^2 \triangleq \mathrm{Y}_2^{i-1}$, $\mathrm{M \triangleq}$ $\mathrm{(M_0,M_1,M_2)}$, $\mathrm{V}_i^1 \triangleq (\mathrm{M},\mathrm{U}_i,\mathrm{K}_i^1)$ and $\mathrm{V}_i^2 \triangleq (\mathrm{M},\mathrm{U}_i,\mathrm{K}_i^2)$. We then start by considering the common rate $R_c$, applying the same steps used in \eqref{Equ:ConvNS1_1}, we have
\begin{align}
R_c &\leq \frac{1}{n} \sum_{i=1}^n \mathbb{I} (\mathrm{U}_i;\mathrm{Z}_i) + \gamma_c(\epsilon_n). 
\label{Equ:Conv1_1}
\end{align}
Next, we consider the confidential rates $(R_0 + R_1)$ intended for the first legitimate receiver. We have
\begin{align}
R_0+R_1 &\overset{(a)}{\leq} \frac{1}{n} \mathrm{\mathbb{I}(M_0 M_1;Y}_1^n \arrowvert \mathrm{M}_2 \mathrm{M}_c) + \gamma_1(\epsilon_n) \nonumber\\
&\leq \frac{1}{n} \mathrm{\mathbb{I}(M;Y}_1^n \arrowvert \mathrm{M}_c) + \gamma_1(\epsilon_n) \nonumber\\
&\overset{(b)}{\leq} \frac{1}{n} \Big[\mathrm{\mathbb{I}(M;Y}_1^n \arrowvert  \mathrm{M}_c) - \mathrm{\mathbb{I}(M;Z}^n \arrowvert \mathrm{M}_c) \Big] + \gamma_1(\epsilon_n,\tau_{n}) \nonumber\\
&= \frac{1}{n} \sum_{i=1}^n \Big[ \mathrm{\mathbb{I}(M;Y}_{1i} \arrowvert \mathrm{M}_c \mathrm{Y}_1^{i-1})-\mathrm{\mathbb{I}(M;Z}_i \arrowvert \mathrm{M}_c \tilde{\mathrm{Z}}^{i+1}) \Big] \nonumber\\
&\qquad+ \gamma_1(\epsilon_n,\tau_{n}) \nonumber\\
&\overset{(c)}{=} \frac{1}{n} \sum_{i=1}^n \Big[ \mathrm{\mathbb{I}(M; Y}_{1i} \arrowvert \mathrm{M}_c \mathrm{Y}_1^{i-1} \tilde{\mathrm{Z}}^{i+1}) \nonumber\\
&\qquad - \mathrm{\mathbb{I}(M;Z}_i \arrowvert \mathrm{M}_c \mathrm{Y}_1^{i-1} \tilde{\mathrm{Z}}^{i+1}) \Big] + \gamma_1(\epsilon_n,\tau_{n}) \nonumber\\
&= \frac{1}{n} \sum_{i=1}^n \Big[\mathbb{I}(\mathrm{V}_i^1; \mathrm{Y}_{1i} \arrowvert \mathrm{U}_i \mathrm{K}_i^1) - \mathbb{I}(\mathrm{V}_i^1;\mathrm{Z}_i \arrowvert \mathrm{U}_i \mathrm{K}_i^1) \Big]\nonumber\\
&\qquad + \gamma_1(\epsilon_n,\tau_n), 
\label{Equ:Conv1_2}
\end{align}
where $(a)$ follows from \eqref{Equ:RelFirCond}; $(b)$ follows from \eqref{Equ:JoiSecCon}, where $\gamma_1(\epsilon_n, \tau_{n}) = \tau_n/n + \gamma_c(\epsilon_n) +\gamma_1(\epsilon_n)$ and $(c)$ follows by the Csisz\'{a}r sum identity \cite[Lemma 7]{Csis}. Following the same steps we can derive a similar bound for the confidential rates $(R_0 + R_2)$ intended for the second legitimate receiver as:
\begin{align}
R_0 + R_2 &\leq \frac{1}{n} \sum_{i=1}^n \Big[\mathbb{I}(\mathrm{V}_i^2; \mathrm{Y}_{2i} \arrowvert \mathrm{U}_i \mathrm{K}_i^2) - \mathbb{I}( \mathrm{V}_i^2;\mathrm{Z}_i \arrowvert \mathrm{U}_i \mathrm{K}_i^2) \Big] \nonumber\\
&\qquad+ \gamma_2(\epsilon_n,\tau_{n}), 
\label{Equ:Conv1_3}
\end{align}
where $\gamma_2(\epsilon_n, \tau_{n}) = \tau_n/n + \gamma_c(\epsilon_n) +\gamma_2(\epsilon_n)$. On the other hand, if we consider the sum of the common rate and the confidential rates $(R_c + R_0 + R_1)$ intended for the first legitimate receiver, we have
\begin{align}
R_c +& R_0 + R_1 \overset{(a)}{\leq} \frac{1}{n} \mathbb{I}(\mathrm{M}_c \mathrm{M_0 M_1;Y}_1^n \arrowvert \mathrm{M}_2) + \tilde{\gamma}_1 (\epsilon_n) \nonumber\\
&\leq \frac{1}{n} \mathbb{I}(\mathrm{M}_c \mathrm{M;Y}_1^n) + \tilde{\gamma}_1 (\epsilon_n) \nonumber\\
&\overset{(b)}{\leq} \frac{1}{n} \Big[ \mathbb{I}(\mathrm{M}_c; \mathrm{Y}_1^n) + \mathbb{I}(\mathrm{M;Y}_1^n \arrowvert \mathrm{M}_c) - \mathrm{\mathbb{I}(M;Z}^n \arrowvert \mathrm{M}_c) \Big]\nonumber\\
&\qquad + \tilde{\gamma}_1(\epsilon_n,\tau_n) \nonumber\\
&\overset{(c)}{\leq} \frac{1}{n} \sum_{i=1}^n \Big[\mathbb{I}(\mathrm{M}_c \tilde{\mathrm{Z}}^{i+1} ;\mathrm{Y}_{1i} \arrowvert \mathrm{Y}_1^{i-1}) + \mathbb{I}(\mathrm{V}_i^1; \mathrm{Y}_{1i} \arrowvert \mathrm{U}_i \mathrm{K}_i^1) \nonumber\\
&\qquad - \mathbb{I}(\mathrm{V}_i^1;\mathrm{Z}_i \arrowvert \mathrm{U}_i \mathrm{K}_i^1) \Big] + \tilde{\gamma}_1(\epsilon_n,\tau_n) \nonumber\\
&= \sum_{i=1}^n \Big[\mathbb{I} (\mathrm{U}_i;\mathrm{Y}_{1i} \arrowvert \mathrm{K}_i^1) + \mathbb{I}(\mathrm{V}_i^1; \mathrm{Y}_{1i} \arrowvert \mathrm{U}_i \mathrm{K}_i^1) \nonumber\\
&\qquad - \mathbb{I}(\mathrm{V}_i^1;\mathrm{Z}_i \arrowvert \mathrm{U}_i \mathrm{K}_i^1) \Big] +  \tilde{\gamma}_1(\epsilon_n,\tau_n) \nonumber\\
&\overset{(d)}{=} \sum_{i=1}^n \Big[\mathbb{I}(\mathrm{V}_i^1; \mathrm{Y}_{1i} \arrowvert \mathrm{K}_i^1) - \mathbb{I}(\mathrm{V}_i^1;\mathrm{Z}_i \arrowvert \mathrm{U}_i \mathrm{K}_i^1) \Big] \nonumber\\
&\qquad +  \tilde{\gamma}_1(\epsilon_n,\tau_n), \label{Equ:Conv1_4}
\end{align}
where $(a)$ follows from \eqref{Equ:RelComCod} and \eqref{Equ:RelFirCond} as $\tilde{\gamma}_1 (\epsilon_n) = \gamma_c(\epsilon_n) + \gamma_1(\epsilon_n)$; $(b)$ follows from \eqref{Equ:JoiSecCon}, where $\tilde{\gamma}_1(\epsilon_n, \tau_{n}) = \tau_n/n + 2\gamma_c (\epsilon_n) +\gamma_1(\epsilon_n)$; $(c)$ follows as in \eqref{Equ:Conv1_2} and the fact that $\mathbb{I}(\mathrm{M}_c \tilde{\mathrm{Z}}^{i+1};\mathrm{Y}_{1i} \arrowvert \mathrm{Y}_1^{i-1}) \geq \mathbb{I}(\mathrm{M}_c;\mathrm{Y}_{1i} \arrowvert \mathrm{Y}_1^{i-1})$; while $(d)$ follows from the chain rule of mutual information. Following the same steps we can derive a similar bound for the sum of the common rate and the confidential rates $(R_c + R_0 + R_2)$ intended for the second legitimate receiver as
\begin{align}
R_c + R_0 + R_2 &\leq \sum_{i=1}^n \Big[\mathbb{I}(\mathrm{V}_i^2; \mathrm{Y}_{2i} \arrowvert \mathrm{K}_i^2) - \mathbb{I}(\mathrm{V}_i^2;\mathrm{Z}_i \arrowvert \mathrm{U}_i \mathrm{K}_i^2) \Big] \nonumber\\
&\qquad+ \tilde{\gamma}_2(\epsilon_n,\tau_n), \label{Equ:Conv1_5}
\end{align}
where $\tilde{\gamma_2}(\epsilon_n,\tau_{n}) = \tau_n/n + 2\gamma_c(\epsilon_n) +\gamma_2(\epsilon_n)$. Now using \eqref{Equ:Conv1_1} - \eqref{Equ:Conv1_5} followed by introducing a random variable $\mathrm{T}$ independent of all others and uniformly distributed over $\llbracket 1; n \rrbracket$ and let $\mathrm{U = (U_T, T)}$, $\mathrm{K^1 = K_T^1}$, $\mathrm{K^2 = K_T^2}$, $\mathrm{V^1 = V_T^1}$, $\mathrm{V^2 = V_T^2}$, $\mathrm{Y_1 = Y_{1T}}$, $\mathrm{Y_2 = Y_{2T}}$ and $\mathrm{Z = Z_T}$, then take the limit as $n \rightarrow \infty$ such that $\gamma_c(\epsilon_n)$, $\gamma_1(\epsilon_n,\tau_{n})$, $\tilde{\gamma_1}(\epsilon_n,\tau_{n})$, $\gamma_2(\epsilon_n, \tau_{n})$ and $\tilde{\gamma}_2 (\epsilon_n,\tau_{n}) \rightarrow 0$, we reach the following
\begin{subequations} \label{Equ:Conv1_6}
\begin{align}
R_c &\leq \mathrm{\mathbb{I} (U;Z)} \label{Equ:Conv1_6_1} \\
R_0 + R_1 &\leq \mathrm{\mathbb{I}(V^1;Y_1 \arrowvert U K^1) - \mathbb{I}(V^1;Z \arrowvert U K^1)} \label{Equ:Conv1_6_2}\\
R_0 + R_2 &\leq \mathrm{\mathbb{I}(V^2;Y_2 \arrowvert U K^2) - \mathbb{I}(V^2;Z \arrowvert U K^2)} \label{Equ:Conv1_6_4} \\
R_c + R_0 + R_1 &\leq \mathrm{\mathbb{I}(V^1;Y_1 \arrowvert  K^1) - \mathbb{I}(V^1;Z \arrowvert U K^1)} \label{Equ:Conv1_6_3}\\
R_c + R_0 + R_2 &\leq \mathrm{\mathbb{I}(V^2;Y_2 \arrowvert  K^2) - \mathbb{I}(V^2;Z \arrowvert U K^2)}, \label{Equ:Conv1_6_5}
\end{align}
\end{subequations}
where $\mathrm{(U,K^1) - V^1 - X - (Y_1,Y_2,Z)}$ and $\mathrm{(U,K^2) - V^2 -}$ $\mathrm{X - (Y_1, Y_2,Z)}$ form Markov chains. First, let us consider \eqref{Equ:Conv1_6_4} and \eqref{Equ:Conv1_6_5}, these two inequalities identify the constraints on the common and confidential rates with respect to the second legitimate receiver, which have an arbitrary relation with the eavesdropper. Since conditional mutual information is the expectation of the unconditional one, Eq. \eqref{Equ:Conv1_6_4} can be upper bounded as follows:
\begin{align}
R_0 +R_2 &\leq \sq{\mathbb{E}_{\mathrm{K}^2} \Big[\mathrm{\mathbb{I}(V^2;Y_2 \arrowvert U,K^2}) - \mathrm{\mathbb{I}(V^2;Z \arrowvert U,K^2}) \Big]} \nonumber\\
&\overset{(a)}{\leq} \mathrm{\mathbb{I}(V^2;Y_2 \arrowvert U,K^2 = }k^{2*}) - \mathrm{\mathbb{I} (V^2;Z \arrowvert U,K^2=}k^{2*}) \nonumber\\
&\overset{(b)}{=} \mathbb{I}(\mathrm{V}^{2*};\mathrm{Y_2 \arrowvert U) - \mathbb{I}(V}^{2*}; \mathrm{Z \arrowvert U)},  \label{Equ:Conv1_7} 
\end{align}
where $\mathrm{U - V^{2*} - X - (Y_1,Y_2,Z)}$ forms a Markov chain. $(a)$ follows because $k^{2*}$ is the value of $\mathrm{K}^2$ such that, $\mathrm{\mathbb{I}(V^2;Y_2 \arrowvert U,K^2 = }k^{2*}) - \mathrm{\mathbb{I}(V^2;Z \arrowvert U,K^2=}k^{2*})$ is greater than or equal $\mathrm{ \mathbb{I} (V^2;Y_2 \arrowvert U,K^2 = }k^2) - \mathrm{\mathbb{I}(V^2;Z \arrowvert U,K^2=}k^2)$, for all $k^2 \in \mathcal{K}^2$; while $(b)$ follows as $\mathrm {V}^{2*}$ is distributed as $\sq{Q(v^2 \arrowvert u,k^2= k^{2*})}$. Similarly we can bound Eq. \eqref{Equ:Conv1_6_5} as follows:
\begin{align}
R_c + R_0 + R_2 &\leq \sq{\mathbb{E}_{\mathrm{K}^2} \Big[\mathrm{\mathbb{I}(V^2;Y_2 \arrowvert K^2}) - \mathrm{\mathbb{I}(V^2;Z \arrowvert U,K^2}) \Big]} \nonumber\\
&\overset{(a)}{\leq} \sq{\mathrm{\mathbb{I}(V^2;Y_2 \arrowvert K^2 = }k^{2\star}) - \mathrm{\mathbb{I} (V^2;Z \arrowvert U,K^2=}k^{2\star})} \nonumber\\
&\overset{(b)}{=} \mathbb{I}(\mathrm{V}^{2\star};\mathrm{Y_2) - \mathbb{I}(V}^{2\star}; \mathrm{Z \arrowvert U)} \nonumber\\
&\overset{(c)}{=} \mathrm{\mathbb{I}(U;Y_2)} + \mathbb{I}(\mathrm{V}^{2\star};\mathrm{Y_2 \arrowvert U) - \mathbb{I}(V}^{2\star}; \mathrm{Z \arrowvert U)} \nonumber\\
&\overset{(d)}{\leq} \mathrm{\mathbb{I}(U;Y_2)} + \mathbb{I}(\mathrm{V}^{2*};\mathrm{Y_2 \arrowvert U) - \mathbb{I}(V}^{2*}; \mathrm{Z \arrowvert U)} \nonumber\\
&= \mathrm{\mathbb{I}(V^{2*};Y_2) - \mathbb{I}(V^{2*};Z \arrowvert U)},
\label{Equ:Conv1_8}
\end{align}
where $(a)$ follows as $k^{2\star}$ is the value of $\mathrm{K}^2$ such that, $\mathrm{\mathbb{I} (V^2;Y_2 \arrowvert K^2 = }k^{2\star}) - \mathrm{\mathbb{I}(V^2;Z \arrowvert U,K^2=}k^{2\star})$ is greater than or equal to $\mathrm{\mathbb{I}(V^2;Y_2 \arrowvert K^2 = }k^2) - \mathrm{\mathbb{I} (V^2;Z \arrowvert U,K^2=}k^2)$, for all $k^2 \in \mathcal{K}^2$; $(b)$ follows as $\mathrm {V}^{2\star}$ is distributed as $\sq{Q(v^2 \arrowvert u,k^2= k^{2\star})}$; $(c)$ follows because $\mathrm{U - V^{2\star} - X - (Y_1,Y_2,Z)}$ forms a Markov chain; and $(d)$ follows because the term $\mathrm{ \mathbb{I}(V^{2\star} ;Y_2 \arrowvert U) - \mathbb{I}(V^{2\star};Z \arrowvert U)}$ can be reformulated as $\mathrm{\mathbb{I}(V^{2} ;Y_2 \arrowvert U,K^2=} k^{2\star}) - \mathrm{\mathbb{I} (V^{2};Z \arrowvert U,K^2=}k^{2\star})$ which is smaller than or equal to $\mathrm{\mathbb{I} (V^{2*};Y_2 \arrowvert U) - \mathbb{I}(V^{2*};Z \arrowvert U)}$. This actually implies that $k^{2*}$ and $k^{2\star}$ are identical and the differences of the mutual information condition on $\mathrm{K}^2$ in \eqref{Equ:Conv1_6_4} and \eqref{Equ:Conv1_6_5} are maximized by the same value. \newline
\indent Now, consider \eqref{Equ:Conv1_6_2} and \eqref{Equ:Conv1_6_3}, these two inequalities identify the constraints on the common and confidential rates with respect to the first legitimate receiver which is more capable than the eavesdropper. In order to simplify these two inequalities we require the following lemma.
\begin{Lemma}
Let $Q(y,z \arrowvert x)$ be a discrete memoryless BC and assume that $\mathrm{Y}$ is more capable than $\mathrm{Z}$. Consider $\mathrm{K}$, $\mathrm{U}^1$, $\mathrm{U}^2$ and $\mathrm{V}$ to be a set of random variables, such that $\mathrm{(U^1,U^2,K) - V - X - (Y,Z)}$ forms a Markov chain. Then the following holds: $\mathrm {\mathbb{I}(V;Y \arrowvert U^1 K) - \mathbb{I}(V;Z \arrowvert U^2 K)} \leq \mathrm {\mathbb{I}(X;Y \arrowvert U^1) - \mathbb{I}(X;Z \arrowvert U^2)}$.
\label{Prop:MoreCap}
\end{Lemma}
\begin{IEEEproof}
This lemma is based on the properties of more capable channels and the definition of the conditional mutual information. A detailed proof is given in Appendix \ref{App:MoreCap} for completeness.
\end{IEEEproof}
$ $\newline
\indent If we apply the previous lemma to Eq. \eqref{Equ:Conv1_6_2} by letting $\mathrm{U^1 = U^2 = U}$, we have
\begin{equation}
R_0 + R_1 \leq \mathrm{\mathbb{I}(X;Y_1 \arrowvert U) - \mathbb{I}(X;Z \arrowvert U)}. \label{Equ:Conv1_9_1}
\end{equation} 
On the other hand if we let $\mathrm{U}^1 = \emptyset$ and $\mathrm{U^2 = U}$, then applied the previous lemma to Eq. \eqref{Equ:Conv1_6_3}, we reach the following bound
\begin{equation}
R_c + R_0 + R_1 \leq \mathrm{\mathbb{I}(X;Y_1) - \mathbb{I}(X;Z \arrowvert U)}. \label{Equ:Conv1_9_2}
\end{equation}
Now, if we combine the upper bounds in \eqref{Equ:Conv1_6_1}, \eqref{Equ:Conv1_7} - \eqref{Equ:Conv1_9_2}, then let $\mathrm{V = V^{2*}}$, such that $\mathrm{U - V - X - (Y_1,Y_2,Z)}$ forms a Markov chain, we reach a region that matches the achievable rate region given by \eqref{Equ:Cap1} and this completes our converse. One last point remains, regarding the cardinality bounds on $\arrowvert \mathcal{U} \arrowvert$ and $\arrowvert \mathcal{V} \arrowvert$, they follow from the Fenchel-Bunt strengthening of the usual Carath\'{e}odory's theorem \cite[Appendix C]{Gamal2}.
\end{IEEEproof}
\begin{Coro}
Consider a wiretap BC with degraded message sets and message cognition where the two legitimate receivers $\mathrm{Y}_1$ and $\mathrm{Y}_2$ are less noisy than the eavesdropper $\mathrm{Z}$, i.e. $\mathrm{Y_1} \succeq \mathrm{Z}$ and $\mathrm{Y_2} \succeq \mathrm{Z}$. Then, the joint secrecy capacity region is given by the set of all rate quadruples $(R_c,R_0,R_1,R_2) \in \mathbb{R}_+^4$ that satisfy
\begin{align}
R_c &\leq \mathrm{\mathbb{I}(U;Z)} \nonumber \\
R_0 + R_1 &\leq \mathrm{\mathbb{I}(X;Y_1\arrowvert U) - \mathbb{I}(X;Z \arrowvert U)} \nonumber \\
R_0 + R_2 &\leq \mathrm{\mathbb{I}(X;Y_2\arrowvert U) - \mathbb{I}(X;Z \arrowvert U)} \nonumber
\end{align}
for some $\mathrm{(U,X)}$, such that $\mathrm{U - X - (Y_1, Y_2, Z)}$ forms a Markov chain. Further it suffices to have $\arrowvert \mathrm{U} \arrowvert \leq \arrowvert \mathrm{X} \arrowvert + 3$.
\end{Coro}
\begin{IEEEproof}
The achievability of the previous region follows as in Theorem~\ref{Ther:1} by substituting $\mathrm{V = X}$, while the converse can be derived using the standard techniques of less noisy channels as in \cite{Gamal}. The previous region was first established in \cite{MansourITW}.  
\end{IEEEproof}
\begin{Coro}
Consider a wiretap BC with message cognition only where the two legitimate receivers $\mathrm{Y}_1$ and $\mathrm{Y}_2$ are more capable than the eavesdropper $\mathrm{Z}$. Then, the joint secrecy capacity region is given by the set of all rate pairs $(R_1,R_2) \in \mathbb{R}_+^2$ that satisfy
\begin{align}
R_1 &\leq \mathrm{\mathbb{I}(X;Y_1) - \mathbb{I}(X;Z)} \nonumber \\
R_2 &\leq \mathrm{\mathbb{I}(X;Y_2) - \mathbb{I}(X;Z)} \nonumber
\end{align}
\end{Coro}
\begin{IEEEproof}
The achievability of the previous region follows as in Theorem~\ref{Ther:1} by substituting $\mathrm{V = X}$ and $\mathrm{U} = \emptyset$, while the converse follows by adapting the more capable condition to the second legitimate receiver $\mathrm{Y}_2$. The previous region was first established in \cite{Mansour}.  
\end{IEEEproof}
\section{The Individual Secrecy Capacity Region}
In this section, we investigate the model of the wiretap BC with degraded message sets and message cognition given by Figure~\ref{Fig:BCWS} under the individual secrecy constraint.
\label{Sec:BCIS}
\subsection{Achievable Rate Region}
\begin{Prop}
An achievable individual secrecy rate region for the wiretap BC with degraded message sets and message cognition is given by the set of all rate quadruples $(R_c,R_0,R_1 = R_{11}+R_{12},R_2= R_{21}+R_{22}) \in \mathbb{R}_+^4$ that satisfy
\begin{align}
R_c &\leq \min \Big [\mathrm{\mathbb{I}(U;Z),\mathbb{I}(U;Y_1),\mathbb{I}(U;Y_2) \Big]} \nonumber \\
\sq{R_{12} = R_{21}} &\leq \sq{\min \Big [R_1, R_2, \mathrm{\mathbb{I}(V_\otimes;Y_1 \arrowvert U) ,\mathbb{I}(V_\otimes;Y_2 \arrowvert U)} \Big ]} \nonumber\\
\sq{R_0 + R_{11}} &\leq \mathrm{\mathbb{I}(V_0 V_1;Y_1 \arrowvert V_\otimes) - \mathbb{I}(V_0 V_1; Z \arrowvert V_\otimes)}  \nonumber \\
\sq{R_0 + R_{22}} &\leq \mathrm{\mathbb{I}(V_0 V_2;Y_2 \arrowvert V_\otimes) - \mathbb{I}(V_0 V_2; Z \arrowvert V_\otimes)}  \nonumber \\
\sq{2R_0 + R_{11} + R_{22}} &\leq \mathrm{\mathbb{I}(V_0 V_1;Y_1 \arrowvert V_\otimes) + \mathbb{I} (V_0 V_2;Y_2 \arrowvert V_\otimes)}\nonumber\\
\mathrm{- \mathbb{I}(V_1;V_2 \arrowvert V_0)} &\mathrm{- \mathbb{I} (V_0V_1V_2;Z \arrowvert V_\otimes) -  \mathbb{I} (V_0;Z \arrowvert V_\otimes)} 
\label{Equ:IRates} 
\end{align}
for random variables with joint probability distribution $Q(u)$ $Q(v_\otimes \arrowvert u)$ $Q(v_0 \arrowvert v_\otimes)$ $Q(v_1,v_2 \arrowvert v_0)$ $Q(x\arrowvert v_1,v_2)$ $Q(y_1, y_2, z \arrowvert x)$, such that $\mathrm{U - V_\otimes - V_0 - (V_1,V_2) - X - (Y_1,Y_2,Z)}$ forms a Markov chain.
\label{Lem:Lemma2}
\end{Prop}
\begin{IEEEproof}
The proof combines the principle of superposition random coding \cite{Csis}, one time pad for Shannon's cipher system \cite{Shann}, the usage of Marton coding for secrecy as in \cite{Gamal}, where strong secrecy is achieved as in \cite{Bloch_Resol,Kram2,MoritzSS}. \vspace{0.2cm} \newline
\underline{\textbf{1. Message sets:}} We consider the following sets: The set of common messages $\mathcal{M}_c = \llbracket 1, 2^{nR_c} \rrbracket$, the set of confidential common messages $\mathcal{M}_0 = \llbracket 1, 2^{nR_0} \rrbracket$, two sets of confidential individual messages $\mathcal{M}_1 = \llbracket 1, 2^{nR_1} \rrbracket$ and $\mathcal{M}_2 = \llbracket 1, 2^{nR_2} \rrbracket$, three sets of randomization messages for secrecy $\mathcal{M}_r =\llbracket 1, 2^{nR_r} \rrbracket$, $\mathcal{M}_{r_1} = \llbracket 1, 2^{nR_{r_1}} \rrbracket$ and $\mathcal{M}_{r_2} = \llbracket 1, 2^{nR_{r_2}} \rrbracket$, finally two additional sets $\mathcal{M}_{t_1} = \llbracket 1, 2^{nR_{t_1}} \rrbracket$ and $\mathcal{M}_{t_2} = \llbracket 1, 2^{nR_{t_2}} \rrbracket$ needed for the construction of Marton coding. Further we divided each confidential individual messages set into two sets as follows: $\mathcal{M}_1 = \mathcal{M}_{11} \times \mathcal{M}_{12}$ and $\mathcal{M}_2 = \mathcal{M}_{21} \times \mathcal{M}_{22}$, where $\mathcal{M}_{11} = \llbracket 1, 2^{nR_{11}} \rrbracket$, $\mathcal{M}_{12} = \llbracket 1, 2^{nR_{12}} \rrbracket$, $\mathcal{M}_{21} = \llbracket 1, 2^{nR_{21}} \rrbracket$ and $\mathcal{M}_{22} = \llbracket 1, 2^{nR_{22}} \rrbracket$. In this division, we force $\mathcal{M}_{12}$ and $\mathcal{M}_{21}$ to be of the same size and use them to construct $\mathcal{M}_\otimes = \llbracket 1, 2^{nR_\otimes} \rrbracket$ by \textit{Xoring} the corresponding elements of both. Additionally we use $\mathcal{M} = \mathcal{M}_0 \times \mathcal{M}_{11} \times \mathcal{M}_{22} \times \mathcal{M}_\otimes$ to abbreviate the modified set of all confidential messages. It is important to note that the message structure forces the following condition:
\begin{equation}
R_\otimes = R_{12} = R_{21} \leq \min \big[ R_1, R_2 \big]
\label{Equ:IRelRate_1}
\end{equation} 

\underline{\textbf{2. Random Codebook $\mathcal{C}^s_n$:}} Fix an input distribution $Q(u,v_\otimes,v_0,v_1,v_2,x)$. Construct the codewords $u^n(m_c)$ for $m_c \in \mathcal{M}_c$ by generating symbols $u_i (m_c)$ with $i \in \llbracket 1, n \rrbracket$ independently according to $Q(u)$. For every $u^n(m_c)$, generate codewords $v_\otimes^n(m_c,m_\otimes)$ for $m_\otimes \in \mathcal{M}_\otimes$ by generating symbols $v_{\otimes_i}(m_c,m_\otimes)$ independently at random according to $Q(v_\otimes \arrowvert u_i(m_c))$. Next, for every $v_\otimes^n (m_c,m_\otimes)$ generate codewords $v_0^n (m_c,m,m_r)$ for $m \in \mathcal{M}$ and $m_r \in \mathcal{M}_r$ by generating symbols $v_{0_i}(m_c,m,m_r)$ independently at random according to $Q(v_0 \arrowvert v_{\otimes_i}(m_c, m_\otimes))$. For each $v_0^n (m_c,m,m_r)$ generate the codewords $v_1^n(m_c,m, m_r,m_{r_1},m_{t_1})$ and $v_2^n(m_c,m,m_r,m_{r_2},m_{t_2})$ for $m_{r_1} \in \mathcal{M}_{r_1}$, $m_{r_2} \in \mathcal{M}_{r_2}$, $m_{t_1} \in \mathcal{M}_{t_1}$ and $m_{t_2} \in \mathcal{M}_{t_2}$ by generating symbols $v_{1_i}(m_c,m,m_r,m_{r_1},m_{t_1})$ and $v_{2_i}(m_c,m,m_r, m_{r_2},m_{t_2})$ independently at random according to $Q(v_1 \arrowvert v_{0_i}(m_c,m,m_r))$ and $Q(v_2 \arrowvert v_{0_i}(m_c,m,m_r))$ respectively. \vspace{0.2cm} 

\underline{\textbf{3. Encoder $E$:}} Given a message pair $(m_c,m)$, where $m = (m_0, m_{11}, m_{22},m_\otimes)$ and $m_\otimes = m_{12} \otimes m_{21}$, the transmitter chooses three randomization messages $m_r$, $m_{r_1}$ and $m_{r_2}$ uniformly at random from the sets $\mathcal{M}_r$, $\mathcal{M}_{r_1}$ and $\mathcal{M}_{r_2}$ respectively. Then, it finds a pair $(m_{t_1},m_{t_2})$ such that $v_1^n(m_c,m,m_r,m_{r_1},m_{t_1})$ and $v_2^n(m_c,m,m_r,m_{r_2},m_{t_2})$ are jointly typical. Finally, it generates a codeword $x^n$ independently at random according to $\prod_{i=1}^n Q (x_i \arrowvert v_{1_i},v_{2_i})$ and transmits it. \vspace{0.2cm} 
 
\underline{\textbf{4. First Legitimate Decoder $\varphi_1$:}} Given $y_1^n$ and its own message $m_2=(m_{21},m_{22})$, outputs $(\hat{m}_c,\hat{m}_0,\hat{m}_1,\hat{m}_r)$; where $\hat{m}_1$ is the concatenation of $\hat{m}_{11}$ and $\hat{m}_{12}$. First it finds the unique messages $(\hat{m}_c, \hat{m}_\otimes,\hat{m},\hat{m}_r)$ such that $u^n(\hat{m}_c)$, $v_\otimes^n (\hat{m}_c, \hat{m}_\otimes)$, $v_0^n(\hat{m}_c,\hat{m},\hat{m_r})$, $v_1^n(\hat{m}_c, \hat{m},\hat{m}_r, \hat{m}_{r_1},\hat{m}_{t_1})$ and $y_1^n$ are jointly typical, for some $(\hat{m}_{r_1}, \hat{m}_{t_1}) \in \mathcal{M}_{r_1} \times \mathcal{M}_{t_1}$, where $\hat{m}= (\hat{m}_0, \hat{m}_{11},m_{22},\hat{m}_\otimes)$. Then, it computes $\hat{m}_{12}$ by \textit{Xoring} $m_{21}$ and $\hat{m}_\otimes$. Otherwise it declares an error. \vspace{0.2cm}

\underline{\textbf{5. Second legitimate Decoder $\varphi_2$:}} Given $y_2^n$ and its own message $m_1=(m_{11},m_{12})$, outputs $(\tilde{m}_c,\tilde{m}_0,\tilde{m}_2,\tilde{m}_r)$; where $\tilde{m}_2$ is the concatenation of $\tilde{m}_{21}$ and $\tilde{m}_{22}$. First it finds the unique messages $(\tilde{m}_c,\tilde{m}_\otimes,\tilde{m},\tilde{m}_r)$ such that $u^n(\tilde{m}_c)$, $v_\otimes^n(\tilde{m}_c,\tilde{m}_\otimes)$, $v_0^n(\tilde{m}_c,\tilde{m}, \tilde{m_r})$, $v_2^n(\tilde {m}_c,\tilde{m},\tilde{m}_r,\tilde{m}_{r_2},\tilde{m}_{t_2})$ and $y_2^n$ are jointly typical, for some $(\tilde{m}_{r_2},\tilde{m}_{t_2}) \in \mathcal{M}_{r_2} \times \mathcal{M}_{t_2}$, where $\tilde{m}= (\tilde{m}_0,m_{11},\tilde{m}_ {22},\tilde{m}_\otimes)$. Then, it computes $\tilde{m}_{21}$ by \textit{Xoring} $m_{12}$ and $\tilde{m}_\otimes$. Otherwise it declares an error.\vspace{0.2cm}

\underline{\textbf{6. Third Eavesdropper Decoder $\varphi_3$:}} Given $z^n$, outputs $\check{m}_c$; if it is the unique message, such that $u^n(\check{m}_c)$ and $z^n$ are jointly typical. Otherwise it declares an error. \vspace{0.2cm}

\underline{\textbf{7. Reliability Analysis:}} We define the error probability of this scheme as
\begin{align*}
\ddot{P}_e (\mathcal{C}_n) &\triangleq \mathbb{P} \big[ (\hat{\mathrm{M}}_c, \mathrm{\hat{M}_\otimes, \hat{M}_0,\hat{M}_{11}},\hat{\mathrm{M}}_r) \neq (\mathrm{M}_c,\mathrm{ M_\otimes,M_0,}\mathrm{M}_{11}, \nonumber\\ 
&\qquad \mathrm{M}_r) \text{ or } (\tilde{\mathrm{M}}_c,\mathrm{ \tilde{M}_\otimes,\tilde{M}_0, \tilde{M}_{22}}, \tilde{\mathrm{M}}_r) \neq (\mathrm{M}_c,\mathrm{M}_\otimes, \nonumber\\
&\qquad \mathrm{M}_0, \mathrm{M}_{22} ,\mathrm{M}_r) \text{ or }  \check{\mathrm{M}}_c \neq \mathrm{M}_c \big].
\end{align*}
We then observe that $\ddot{P}_e (\mathcal{C}_n) \geq P_e (\mathcal{C}_n)$, cf. \eqref{Equ:ErrorProb}. Using the standard analysis of random coding we can prove that for a sufficiently large $n$, with high probability $\ddot{P}_e (\mathcal{C}_n) \leq \epsilon_n$ if
\begin{gather}
R_c \leq \min \Big [\mathrm{\mathbb{I}(U;Y_1), \mathbb{I}(U;Y_2), \mathbb{I}(U;Z)} \Big] - \delta_n(\epsilon_n) \nonumber \\ 
R_\otimes \leq \min \Big [\mathrm{\mathbb{I}(V_\otimes;Y_1 \arrowvert U), \mathbb{I}(V_\otimes;Y_2 \arrowvert U)} \Big]-\delta_n(\epsilon_n) \nonumber \\  
R_{t_1} + R_{t_2} \geq \mathrm{\mathbb{I} (V_1;V_2 \arrowvert V_0)} +  \delta_n (\epsilon_n) \nonumber \\ 
R_0 + R_{11} + R_r + R_{r_1} + R_{t_1} \leq \mathrm{ \mathbb{I} (V_0 V_1; Y_1 \arrowvert V_\otimes)} - \delta_n (\epsilon_n) \nonumber \\ 
R_0 + R_{22} + R_r + R_{r_2} + R_{t_2} \leq \mathrm{ \mathbb{I} (V_0 V_2; Y_2 \arrowvert V_\otimes)} - \delta_n (\epsilon_n).  \label{Equ:IRelRate_2}
\end{gather}

\underline{\textbf{8. Secrecy Analysis:}} Because of the new message sets structure, the random variable $\mathrm{M}_1$ is identified as the product of two independent and uniformly distributed random variables $\mathrm{M}_{11}$ and $\mathrm{M}_{12}$. This also applies to $\mathrm{M}_2$ which is the product of two independent and uniformly distributed random variables $\mathrm{M}_{21}$ and $\mathrm{M}_{22}$. Thus, the individual secrecy constraint given by \eqref{Equ:IndCond} becomes 
\begin{align}
\mathbb{I}\mathrm{(M_0 M_{11} ;Z}^n) + \mathbb{I}\mathrm{ (M_{12} ;Z}^n \arrowvert \mathrm{M_0 M_{11}}) &\leq \tau_{1n} \nonumber\\
\mathbb{I}\mathrm{(M_0 M_{22} ;Z}^n) + \mathbb{I}\mathrm{ (M_{21} ;Z}^n \arrowvert \mathrm{M_0 M_{22}}) &\leq \tau_{2n}.
\label{Equ:LekSum_1}
\end{align}
The term $\mathbb{I}(\mathrm{M_{12};Z}^n \arrowvert \mathrm{M_0 M_{11}})$ represents the leakage of $\mathrm{M}_ {12}$ to the eavesdropper given $\mathrm{M}_0$ and $\mathrm{M}_{11}$. One can proof that this term vanishes as
\begin{align}
\mathbb{I}(\mathrm{M_{12};Z}^n \arrowvert \mathrm{M_0 M_{11}}) &=  \mathrm{\mathbb{H}(M_{12}\arrowvert M_0 M_{11}) - \mathbb{H} (M_{12}} \arrowvert \mathrm{Z}^n \mathrm{M_0 M_{11}}) \nonumber\\
&\overset{(a)}{=} \mathrm{\mathbb{H}(M_{12}) - \mathbb{H} (M_{12}} \arrowvert \mathrm{Z}^n \mathrm{M_0 M_{11}}) \nonumber\\
&\overset{(b)}{\leq} \mathrm{\mathbb{H}(M_{12}) - \mathbb{H}(M_{12} \arrowvert M_\otimes}) \overset{(c)}{=} 0,  \label{Equ:ShanSec_1}
\end{align}
where $(a)$ follows because $\mathrm{M_{12}}$, $\mathrm{M}_0$ and $\mathrm{M_ {11}}$ are independent; $(b)$ follows because the best the eavesdropper can do is to decode $\mathrm{M}_ \otimes$; while $(c)$ follows because of the principle of one time pad in Shannon's cipher system where the entropy of the secret key $\mathbb{H} (\mathrm{M}_{21})$ is equal to the entropy of the transmitted message $\mathbb{H}(\mathrm{M}_{12})$. Using the same steps, we can prove that the term $\mathbb{I}\mathrm{ (M_{21} ;Z}^n \arrowvert \mathrm{M_0 M_{22}})$ which represents the leakage of $\mathrm{M}_ {21}$ to the eavesdropper given $\mathrm{M}_0$ and $\mathrm{M}_{22}$ also vanishes. On the other hand, for a sufficiently large $n$ and $\tau_n \geq \max(\tau_{1n},\tau_{2n}) >0$, the terms $\mathbb{I}\mathrm{(M_0 M_{11} ;Z}^n)$ and $\mathbb{I} \mathrm{(M_0 M_{22} ;Z}^n)$ are with high probability smaller than $\tau_n$, if
\begin{align}
R_r &\geq \mathrm{\mathbb{I}(V_0;Z \arrowvert V_\otimes)} + \delta_n(\tau_n) \nonumber \\
R_r + R_{r_1} + R_{t_1} &\geq \mathrm{\mathbb{I}(V_0 V_1;Z \arrowvert V_\otimes)} + \delta_n (\tau_n) \nonumber \\
R_r + R_{r_2} + R_{t_2} &\geq \mathrm{\mathbb{I}(V_0 V_2;Z \arrowvert V_\otimes)} + \delta_n (\tau_n) \nonumber \\
R_r + R_{r_1} + R_{r_2} &\geq  \mathrm{\mathbb{I}(V_0 V_1 V_2;Z \arrowvert V_\otimes)} +\delta_n(\tau_n).
\label{Equ:ISecRate}
\end{align}
This follows from applying the strong secrecy approach in \cite{Kram2} to the codebook structure of the individual secrecy as we did for the joint secrecy in the previous section. This implies that under the previous constraints, the leakage terms in \eqref{Equ:LekSum_1} are with high probability smaller than $\tau_n$. \vspace{0.2cm} \newline
\indent Now, if we combine \eqref{Equ:IRelRate_1}, \eqref{Equ:IRelRate_2} and \eqref{Equ:ISecRate}, then take the limit as $n \rightarrow \infty$, which implies that $\delta_n(\epsilon_n)$ and $\delta_n (\tau_n) \rightarrow 0$, we prove the achievability of any rate quadruple $(R_c,R_0,R_1,R_2)$ satisfying \eqref{Equ:IRates}.
\end{IEEEproof}

\subsection{Secrecy Capacity For A Class of More Capable Channels}
\begin{mytheo}
Consider a wiretap BC with degraded message sets and message cognition, where the two legitimate receivers $\mathrm{Y_1}$ and $\mathrm{Y_2}$ are more capable than the eavesdropper $\mathrm{Z}$. Then, the individual secrecy capacity region is given by the set of all rate quadruples $(R_c,R_0,R_1,R_2) \in \mathbb{R}_+^4$ that satisfy
\begin{align}
R_c &\leq \mathrm{\mathbb{I}(U;Z)} \nonumber \\
R_0 + R_1 &\leq \mathrm{\mathbb{I}(X;Y_1\arrowvert U) - \mathbb{I}(X;Z \arrowvert U)} + R_\otimes \nonumber \\
R_0 + R_2 &\leq \mathrm{\mathbb{I}(X;Y_2\arrowvert U) - \mathbb{I}(X;Z \arrowvert U)} + R_\otimes \nonumber \\
R_c + R_0 + R_1 &\leq \mathrm{\mathbb{I}(X;Y_1) - \mathbb{I}(X;Z \arrowvert U)} + R_\otimes  \nonumber \\
R_c + R_0 + R_2 &\leq \mathrm{\mathbb{I}(X;Y_2) - \mathbb{I}(X;Z \arrowvert U)} + R_\otimes
\label{Equ:ICap1} 
\end{align}
where $R_\otimes = \min\Big [R_1,R_2,\mathrm{\mathbb{I}(X;Z \arrowvert U)} \Big]$, for some $\mathrm{(U,X)}$, such that $\mathrm{U - X - (Y_1, Y_2, Z)}$ forms a Markov chain. Further it suffices to have $\arrowvert \mathcal{U} \arrowvert \leq \arrowvert \mathcal{X} \arrowvert + 2$.
\end{mytheo}
\begin{IEEEproof}
The achievability is based on the same principle used in Proposition~\ref{Lem:Lemma2}. We start by modifying the structure of the random codebook as follows: For every $u^n(m_c)$, we generate the codewords $x^n(m_c,m,m_r)$ by generating symbols $x_i(m_c,m,m_r)$ independently at random according to $Q(x \arrowvert u_i(m_c))$. Given a message pair $(m_c,m)$, the encoder chooses a message $m_r$ uniformly at random from the set $\mathcal{M}_r$ and transmits $x^n(m_c,m,m_r)$. This changes the decoder at the first legitimate receiver such that, it outputs $(\hat{m}_c,\hat{m},\hat{m}_r)$, if it is the unique triple, where $u^n(\hat{m}_c)$, $x^n(\hat{m}_c,\hat{m},\hat{m}_r)$ and $y_1^n$ are jointly typical. The decoder at the second legitimate receiver also changes in the same way, while the decoder as the eavesdropper is kept unchanged. Under these modifications, the reliability conditions in \eqref{Equ:IRelRate_2} changes to: 
\begin{align}
R_c &\leq \mathrm{\mathbb{I}(U;Z)} -\delta_n(\epsilon_n) \nonumber\\
R_0 + R_{11} + R_\otimes + R_r &\leq \mathrm{\mathbb{I}(X;Y_1 \arrowvert U)} - \delta_n(\epsilon_n) \nonumber\\
R_0 + R_{22} + R_\otimes + R_r &\leq \mathrm{\mathbb{I}(X;Y_2 \arrowvert U)} - \delta_n(\epsilon_n) \nonumber\\ 
R_c + R_0 + R_{11} + R_\otimes + R_r &\leq \mathrm{\mathbb{I}(X;Y_1)} - \delta_n(\epsilon_n) \nonumber\\
R_c + R_0 + R_{22} + R_\otimes + R_r &\leq \mathrm{\mathbb{I}(X;Y_2)} - \delta_n(\epsilon_n).
\label{Equ:ICapLB1_1}
\end{align}    
On the other hand, the secrecy conditions in \eqref{Equ:ISecRate} simplifies to
\begin{equation}
R_\otimes + R_r  \geq  \mathrm{\mathbb{I}(X;Z \arrowvert U)} +\delta_n(\tau_n).
\label{Equ:ICapLB1_2}
\end{equation}
Now using Fourier-Motzkin elimination on the rate constraints given in \eqref{Equ:IRelRate_1},  \eqref{Equ:ICapLB1_1} and \eqref{Equ:ICapLB1_2} followed by taking the limit as $n \rightarrow \infty$, which implies that $\delta_n(\epsilon_n)$ and $\delta_n(\tau_n) \rightarrow 0$, leads the achievability of any rate quadruple $(R_c,R_0,R_1,R_2)$ satisfying \eqref{Equ:ICap1}. \newline
\indent Before jumping to the converse, it is important to highlight the difference between the coding structure in Proposition~\ref{Lem:Lemma2} and in this theorem. In this theorem, the two secrecy encoding techniques: one time pad secret key encoding and wiretap random coding were combined in the same layer $\mathrm{X}$. On the other hand, in Proposition~\ref{Lem:Lemma2} they were structured into two different layers. In the first layer $\mathrm{V}_\otimes$ was used for the one time pad secret key encoding, while the wiretap random coding was performed in the next layers using $\mathrm{V_0 \text{, } V_1 \text{ and } V_2}$. This is because combining the two techniques in the same layer is only possible, if the two legitimate receivers have a statistical advantage over the eavesdropper, such that $\mathbb{I} \mathrm{(X;Y_1)}$ and $\mathbb{I}\mathrm{(X;Y_2)}$ are greater than $\mathbb{I} \mathrm{(X;Z)}$. Otherwise, the conditions in \eqref{Equ:ICapLB1_1} and \eqref{Equ:ICapLB1_2} can not be fulfilled simultaneously leading to a decoding failure. \newline
\indent Now for the converse, we start by letting $\mathrm{U}_i \triangleq (\mathrm{M}_c,\tilde {\mathrm{Z}}^{i+1})$, $\mathrm{K}_i^1 \triangleq \mathrm{Y}_1^{i-1}$, $\mathrm{K}_i^2 \triangleq \mathrm{Y}_2^{i-1}$, $\mathrm{M \triangleq (M_0,M_1,M_2)}$, $\mathrm{V}_i^1 \triangleq (\mathrm{M}, \mathrm{U}_i,\mathrm{K}_i^1)$ and $\mathrm{V}_i^2 \triangleq (\mathrm{M},\mathrm{U}_i, \mathrm{K}_i^2)$. Using the same steps carried out in \eqref{Equ:ConvNS1_1}, we have
\begin{equation}
R_c \leq  \frac{1}{n} \sum_{i=1}^n \mathbb{I} (\mathrm{U}_i;\mathrm{Z}_i) + \gamma_c(\epsilon_n). 
\label{Equ:IConv1_1}
\end{equation}
Next, let us consider the confidential rates $(R_0 + R_1)$ intended to the first legitimate receiver, we have
\begin{align}
R_0 &+ R_1 \overset{(a)}{\leq} \frac{1}{n} \mathrm{\mathbb{I}(M_0 M_1;Y}_1^n \arrowvert \mathrm{M}_2 \mathrm{M}_c) + \gamma_1 (\epsilon_n) \nonumber\\
&\leq \frac{1}{n} \mathrm{\mathbb{I}(M;Y}_1^n \arrowvert \mathrm{M}_c) + \gamma_1(\epsilon_n)  \nonumber\\
&\overset{(b)}{\leq} \frac{1}{n} \Big[ \mathrm{\mathbb{I}(M;Y}_1^n \arrowvert \mathrm{M}_c) - \max \big[ \mathrm{\mathbb{I}(M_0 M_1;Z}^n \arrowvert \mathrm{M}_c),\nonumber\\
&\quad \quad \mathrm{\mathbb{I}(M_0 M_2;Z}^n \arrowvert \mathrm{M}_c) \big] \Big] + \gamma_1 (\epsilon_n,\tau_n)  \nonumber\\
&= \frac{1}{n} \Big[ \mathrm{\mathbb{I}(M;Y}_1^n \arrowvert \mathrm{M}_c) - \mathrm{\mathbb{I}(M; Z}^n \arrowvert \mathrm{M}_c)\Big] + \gamma_1(\epsilon_n,\tau_n) \nonumber\\
&\quad +\frac{1}{n} \min \Big[\mathrm{\mathbb{I}(M_1;Z}^n \arrowvert \mathrm{M_0M_2M}_c), \mathrm{ \mathbb{I}(M_2;Z}^n \arrowvert \mathrm{M_0 M_1 M}_c) \Big] \nonumber\\
&\overset{(c)}{\leq} \frac{1}{n} \Big[ \mathrm{\mathbb{I}(M;Y}_1^n \arrowvert \mathrm{M}_c) - \mathrm{\mathbb{I}(M;Z}^n \arrowvert \mathrm{M}_c) \Big] + \min \big[R_1, R_2\big] \nonumber\\
&\quad \quad + \gamma_1(\epsilon_n,\tau_n) \nonumber\\
&\overset{(d)}{=} \frac{1}{n} \sum_{i=1}^n \Big[\mathbb{I}(\mathrm{V}_i^1; \mathrm{Y}_{1i} \arrowvert \mathrm{U}_i \mathrm{K}_i^1) - \mathbb{I}(\mathrm{V}_i^1;\mathrm{Z}_i \arrowvert \mathrm{U}_i \mathrm{K}_i^1) \Big] \nonumber\\
&\quad \quad + \min \big[R_1, R_2\big] + \gamma_1(\epsilon_n,\tau_n), 
\label{Equ:IConv1_2}
\end{align}
where $(a)$ follows from \eqref{Equ:RelFirCond}; $(b)$ follows after modifying the individual secrecy conditions in \eqref{Equ:IndCond} to include the conditioning on the common message $\mathrm{M}_c$ based on \eqref{Equ:RelComCod} and Lemma~\ref{Prop:CMC}; $(c)$ follows because of the fact that $R_1 \geq \mathrm{\mathbb{I}(M_1;Z}^n \arrowvert \mathrm{M_0 M_2 M}_c)$ and $ R_2 \geq \mathrm{\mathbb{I} (M_2;Z}^n \arrowvert \mathrm{M_0 M_1 M}_c)$; and $(d)$ follows as in \eqref{Equ:Conv1_2}. Following the same steps we can derive a similar bound for the confidential rates $(R_0+R_2)$ intended to the second legitimate receiver as: 
\begin{align}
R_0 + R_2 &\leq \frac{1}{n} \sum_{i=1}^n \Big[\mathbb{I}(\mathrm{V}_i^2; \mathrm{Y}_{2i} \arrowvert \mathrm{U}_i \mathrm{K}_i^2) - \mathbb{I}(\mathrm{V}_i^2;\mathrm{Z}_i \arrowvert \mathrm{U}_i \mathrm{K}_i^2) \Big] \nonumber\\
&\quad \quad +\min \big[R_1, R_2\big] + \gamma_2(\epsilon_n,\tau_n). \label{Equ:IConv1_3}
\end{align}
Now, consider the sum of the common rate and the confidential rates to the first and the second legitimate receivers. Using the techniques used in \eqref{Equ:IConv1_2} and \eqref{Equ:Conv1_4}, we can derive the following bounds
\begin{align}
R_c + R_0 + R_1 &\leq \frac{1}{n} \sum_{i=1}^n \Big[\mathbb{I}(\mathrm{V}_i^1; \mathrm{Y}_{1i} \arrowvert \mathrm{K}_i^1) - \mathbb{I}(\mathrm{V}_i^1;\mathrm{Z}_i \arrowvert \mathrm{U}_i \mathrm{K}_i^1) \Big] \nonumber\\
&\quad \quad + \min \big[R_1, R_2\big] + \tilde{\gamma}_1(\epsilon_n,\tau_n) \nonumber\\
R_c + R_0 + R_2 &\leq \frac{1}{n} \sum_{i=1}^n \Big[\mathbb{I}(\mathrm{V}_i^2; \mathrm{Y}_{2i} \arrowvert \mathrm{K}_i^2) - \mathbb{I}(\mathrm{V}_i^2;\mathrm{Z}_i \arrowvert \mathrm{U}_i \mathrm{K}_i^2) \Big] \nonumber\\
&\quad \quad + \min \big[R_1, R_2\big] + \tilde{\gamma}_2(\epsilon_n,\tau_n).
\label{Equ:IConv1_4}
\end{align}
Now using \eqref{Equ:IConv1_1}, \eqref{Equ:IConv1_2}, \eqref{Equ:IConv1_3} and \eqref{Equ:IConv1_4}, followed by introducing a random variable $\mathrm{T}$ independent of all others and uniformly distributed over $\llbracket 1; n \rrbracket$ and let $\mathrm{U = (U_T,T)}$, $\mathrm{K^1 = K_T^1}$, $\mathrm{K^2 = K_T^2}$, $\mathrm{V^1 = V_T^1}$, $\mathrm{V^2 = V_T^2}$, $\mathrm{Y_1 = Y_{1T}}$, $\mathrm{Y_2 = Y_{2T}}$ and $\mathrm{Z = Z_T}$, then take the limit as $n \rightarrow \infty$ such that $\gamma_c(\epsilon_n)$, $\gamma_1(\epsilon_n,\tau_{n})$, $\tilde{\gamma_1}(\epsilon_n,\tau_{n})$, $\gamma_2(\epsilon_n, \tau_{n})$ and $\tilde{\gamma}_2 (\epsilon_n,\tau_{n}) \rightarrow 0$, we reach the following
\begin{align*}
R_c &\leq \mathrm{\mathbb{I} (U;Z)} \nonumber \\
\sq{R_0 + R_1} &\leq \sq{\mathrm{\mathbb{I}(V^1;Y_1 \arrowvert U K^1) - \mathbb{I}(V^1;Z \arrowvert U K^1)} + \min \big[ R_1, R_2 \big]} \nonumber \\
\sq{R_0 + R_2} &\leq \sq{\mathrm{\mathbb{I}(V^2;Y_2 \arrowvert U K^2) - \mathbb{I}(V^2;Z \arrowvert U K^2)} + \min \big[ R_1, R_2 \big]} \nonumber \\
\sq{R_c + R_0 + R_1} &\leq \sq{\mathrm{\mathbb{I}(V^1;Y_1 \arrowvert  K^1) - \mathbb{I}(V^1;Z \arrowvert U K^1)} + \min \big[ R_1, R_2 \big]} \nonumber \\
\sq{R_c + R_0 + R_2} &\leq \sq{\mathrm{\mathbb{I}(V^2;Y_2 \arrowvert  K^2) - \mathbb{I}(V^2;Z \arrowvert U K^2)} + \min \big[ R_1, R_2 \big]}.
\end{align*}
Since $\mathrm{Y}_1$ and $\mathrm{Y}_2$ are more capable than $\mathrm{Z}$, we can use Lemma~\ref{Prop:MoreCap} to modify the previous bounds as we did in modifying the bounds in \eqref{Equ:Conv1_6_2} and \eqref{Equ:Conv1_6_3} to \eqref{Equ:Conv1_9_1} and \eqref{Equ:Conv1_9_2}  as follows:
\begin{align}
R_c &\leq \mathrm{\mathbb{I} (U;Z)} \nonumber \\
R_0 + R_1 &\leq \mathrm{\mathbb{I}(X;Y_1 \arrowvert U) - \mathbb{I}(X;Z \arrowvert U)} 
+ \min \big[ R_1, R_2 \big] \nonumber \\
R_0 + R_2 &\leq \mathrm{\mathbb{I}(X;Y_2 \arrowvert U) - \mathbb{I}(X;Z \arrowvert U)} 
+ \min \big[ R_1, R_2 \big]\nonumber \\
\sq{R_c + R_0 + R_1} &\leq \mathrm{\mathbb{I}(X;Y_1) - \mathbb{I}(X;Z \arrowvert U)} 
+ \min \big[ R_1, R_2 \big] \nonumber \\
\sq{R_c + R_0 + R_2} &\leq \mathrm{\mathbb{I}(X;Y_2) - \mathbb{I}(X;Z \arrowvert U)} 
+ \min \big[ R_1, R_2 \big].
\label{Equ:IConv1_6}
\end{align} 
To finalize our converse, we need to highlight the upper bounds required for reliable communication established in \eqref{Equ:ConvNS1_7} and \eqref{Equ:ConvNS1_8} in addition to the standard upper bound in \eqref{Equ:ConvNS1_9}. These bounds will impose and additional constraint to guarantee that the addition of the minimum of $R_1$ and $R_2$ does not contradict them. Thus the term $\min [R_1, R_2]$ will change to $\min [R_1,R_2,\mathrm{\mathbb{I}(X;Z \arrowvert U)}]$. Introducing this modification to the bounds in \eqref{Equ:IConv1_6} matches the rate region in \eqref{Equ:ICap1} and this completes our converse.
\end{IEEEproof}
\begin{Coro}
Consider a wiretap BC with degraded message sets and message cognition where the two legitimate receivers $\mathrm{Y}_1$ and $\mathrm{Y}_2$ are less noisy than the eavesdropper $\mathrm{Z}$, i.e. $\mathrm{Y_1} \succeq \mathrm{Z}$ and $\mathrm{Y_2} \succeq \mathrm{Z}$. Then, the individual secrecy capacity region is given by the set of all rate quadruples $(R_c,R_0,R_1,R_2) \in \mathbb{R}_+^4$ that satisfy
\begin{align}
R_c &\leq \mathrm{\mathbb{I}(U;Z)} \nonumber \\
\sq{R_0 + R_1} &\leq \sq{\mathrm{\mathbb{I}(X;Y_1\arrowvert U) - \mathbb{I}(X;Z \arrowvert U)} + \min\Big [R_1,R_2,\mathrm{\mathbb{I}(X;Z \arrowvert U)} \Big]}  \nonumber\\
\sq{R_0 + R_2} &\leq \sq{\mathrm{\mathbb{I}(X;Y_2\arrowvert U) - \mathbb{I}(X;Z \arrowvert U)} + \min\Big [R_1,R_2,\mathrm{\mathbb{I}(X;Z \arrowvert U)} \Big]}  \nonumber 
\end{align}
for some $\mathrm{(U,X)}$, such that $\mathrm{U - X - (Y_1, Y_2, Z)}$ forms a Markov chain. Further it suffices to have $\arrowvert \mathrm{U} \arrowvert \leq \arrowvert \mathrm{X} \arrowvert + 3$.
\end{Coro}
\begin{IEEEproof}
The previous region is a special case from the region in Theorem~\ref{Ther:2}, where the sum rate bound is not needed because of the properties of less noisy channels. The previous region was first established in \cite{MansourITW}.  
\end{IEEEproof}

\section{Conclusion}
We studied a three-receiver broadcast channel with degraded message sets and message cognition. We established the non-secrecy capacity region for the general case by providing a weak converse showing that the straightforward extension of the K\"{o}rner and Marton bound to our model is optimal. We then investigated, evaluated and compared the performance of two different secrecy constraints for our model: the joint secrecy and the individual one. For each constraint we derived a general achievable rate region. We further showed that the principle of indirect decoding is optimal for the joint secrecy criterion, if only one of the legitimate receivers is more capable than the eavesdropper, such that it establishes the capacity. On the other hand, we managed to establish the individual secrecy capacity if the two receivers are more capable than the eavesdropper. Our results indicate that the individual secrecy provides a larger capacity region as compared to the joint one. This increase arises from the mutual trust between the legitimate receivers in the individual secrecy constraint that allows the usage of secret key encoding, which is not possible for the conservative joint secrecy.


%

\appendix
\subsection{Proof of Lemma \ref{Prop:LessN}}
\label{App:LessN}
\begin{IEEEproof}
We define $\Delta = \frac{1}{n} [\mathrm{\mathbb{I}(M;Z}^n \arrowvert \mathrm{W}) -\mathrm{ \mathbb{I}(M;Y}^n \arrowvert \mathrm{W})]$ and prove that if $\mathrm{Y \succeq Z}$, $\Delta \leq 0$ and this directly implies our proposition. Let $\mathrm{U}_i \triangleq ( \mathrm{W}, \tilde{\mathrm{Z}} ^{i+1}, \mathrm{Y}^{i-1})$ and $\mathrm{V}_i \triangleq (\mathrm{M}, \mathrm{U}_i)$, we have
\begin{align}
\Delta &= \frac{1}{n} \sum_{i=1}^n \Big[ \mathrm{\mathbb{I}(M;Z}_i \arrowvert \mathrm{W} \tilde{\mathrm{Z}}^{i+1}) - \mathrm{\mathbb{I}(M;Y}_i \arrowvert \mathrm{W}\mathrm{Y}^{i-1}) \Big] \nonumber\\
&\overset{(a)}{=} \frac{1}{n} \sum_{i=1}^n \Big[ \mathrm{\mathbb{I}(M;Z}_i \arrowvert \mathrm{W}\tilde{\mathrm{Z}}^{i+1} \mathrm{Y}^{i-1}) - \mathrm{\mathbb{I}(M;Y}_i \arrowvert \mathrm{W}\tilde{\mathrm{Z}}^{i+1} \mathrm{Y}^{i-1}) \Big] \nonumber \\
&= \frac{1}{n} \sum_{i=1}^n \Big[\mathbb{I}(\mathrm{V}_i; \mathrm{Z}_i \arrowvert \mathrm{U}_i) - \mathbb{I}(\mathrm{V}_i; \mathrm{Y}_i \arrowvert \mathrm{U}_i)\Big] \nonumber \\
&\overset{(b)}{=} \mathrm{\mathbb{I}(V;Z \arrowvert U) - \mathbb{I}(V;Y \arrowvert U)} \nonumber \\
&= \mathbb{E}_\mathrm{U} \Big [\mathrm{\mathbb{I}(V;Z \arrowvert U}) - \mathrm{ \mathbb{I}(V;Y \arrowvert U}) \Big]\nonumber \\
&\overset{(c)} {\leq} \mathrm{\mathbb{I}(V;Z \arrowvert  U}=u^*) - \mathrm{ \mathbb{I} (V;Y \arrowvert U}=u^*) \nonumber\\
&\overset{(d)}{\leq} \mathrm{\mathbb{I}(V^*;Z) - \mathbb{I}(V^*;Y)} \overset{(e)} {\leq} 0
\label{Equ:LNCon} 
\end{align}
where $(a)$ follows from the Csisz\'{a}r sum identity \cite[Lemma 7]{Csis}; $(b)$ follows by introducing a random variable $\mathrm{T}$ independent of all others and uniformly distributed over $\llbracket 1,n \rrbracket$, then  letting $\mathrm{U =(U}_T ,\mathrm{T})$, $\mathrm{V =V}_T$, $\mathrm{Y}=\mathrm{Y}_{T}$ and $\mathrm{Z=Z}_T$; $(c)$ follows as $u^*$ is the value of $\mathrm{U}$ that maximizes the difference; $(d)$ follows as $\mathrm{V}^*$ is distributed as $Q(v \arrowvert u=u^*)$ cf. \cite[Corollary 2.3]{LiangITS} and $(e)$ follows since $\mathrm{V^* - X - (Y,Z)}$ forms a Markov chain and $\mathrm{Y \succeq Z}$.
\end{IEEEproof}

\subsection{Proof of Lemma \ref{Prop:CMC}}
\label{App:CMC}
\begin{IEEEproof}
We have
\begin{align*}
\mathbb{I}(\mathrm{M};\mathrm{Z}^n \arrowvert \mathrm{W}) &= \mathbb{H}( \mathrm{M} \arrowvert \mathrm{W}) - \mathbb{H}(\mathrm{M} \arrowvert \mathrm{Z}^n \mathrm{W}) \nonumber \\
&\leq \mathbb{H}( \mathrm{M} \arrowvert \mathrm{W}) - \mathbb{H}(\mathrm{M} \arrowvert \mathrm{Z}^n \mathrm{W}) - \mathbb{H}( \mathrm{W} \arrowvert \mathrm{Z}^n) + \alpha \nonumber \\
&\overset{(a)}{=} \mathbb{H}( \mathrm{M}) - \mathbb{H}(\mathrm{W} \mathrm{M} \arrowvert \mathrm{Z}^n ) + \alpha \nonumber \\
&= \mathbb{H}( \mathrm{M}) - \mathbb{H}( \mathrm{M} \arrowvert \mathrm{Z}^n) - \mathbb{H}(\mathrm{W} \arrowvert \mathrm{Z}^n \mathrm{M}) + \alpha \nonumber \\
&\overset{(b)}{\leq} \mathbb{I}(\mathrm{M};\mathrm{Z}^n) + \alpha \leq \alpha + \beta,
\end{align*}
where $(a)$ follows because M and W are independent, while $(b)$ follows because $\mathbb{H}(\mathrm{W} \arrowvert \mathrm{Z}^n \mathrm{M}) \geq 0$. 
\end{IEEEproof}

\subsection{Proof of Lemma \ref{Prop:MoreCap}}
\label{App:MoreCap}
\begin{IEEEproof}
Let $\Theta \triangleq \mathrm {\mathbb{I}(V;Y \arrowvert U^1 K) - \mathbb{I}(V;Z \arrowvert U^2 K)}$, we have
\begin{align*}
\Theta &= \mathbb{E}_{\mathrm{K}} \Big[ \mathrm{\mathbb{I}(V;Y \arrowvert U^1,K)} - \mathrm{ \mathbb{I}(V;Z \arrowvert U^2,K)} \Big]\nonumber \\ 
&\overset{(a)} {\leq} \mathrm{\mathbb{I}(V;Y \arrowvert U^1,K=}k^*) - \mathrm{ \mathbb{I}(V;Z \arrowvert U^2,K=}k^*) \nonumber\\
&\overset{(b)}{=} \mathrm{\mathbb{I}(V^*;Y \arrowvert U^1) - \mathbb{I}(V^*;Z \arrowvert U^2)} \nonumber\\
&\overset{(c)}{=} \mathrm{\mathbb{I}(X;Y \arrowvert U^1) - \mathbb{I}(X;Y \arrowvert V^*) - \mathbb{I}(X;Z \arrowvert U^2) + \mathbb{I}(X;Z \arrowvert V^*)} \nonumber\\
&\overset{(d)}{\leq} \mathrm{\mathbb{I}(X;Y \arrowvert U^1) - \mathbb{I}(X;Z \arrowvert U^2)}, 
\end{align*}
where $(a)$ follows as $k^*$ is the value of $\mathrm{K}$ that maximizes the difference; $(b)$ follows as $\mathrm{V}^*$ is distributed as $Q (v \arrowvert u^1, u^2, k=k^*)$; $(c)$ follows since $\mathrm{(U^1,U^2) - V^* - X - (Y,Z)}$ forms a Markov chain and $(d)$ follows because $\mathrm{Y}$ is more capable than $\mathrm{Z}$, which implies that $\mathrm{\mathbb{I}(X;Y \arrowvert V^*)} \geq \mathrm{\mathbb{I}(X;Z \arrowvert V^*)}$.
\end{IEEEproof}

\ifCLASSOPTIONcaptionsoff
  \newpage
\fi

\bibliographystyle{IEEEtran}
\bibliography{IEEEabrv,biblography}

\end{document}